\documentclass[ reprint, aps,
]{revtex4-2}
\usepackage{amsmath}
\usepackage{graphicx}
\usepackage{dcolumn}
\usepackage{bm}
\usepackage{caption}
\usepackage{subcaption}
\usepackage{multirow}
\usepackage{hyperref}
\begin{document}

\title{Cosmo-dynamics of dark energy models resulting from a parametrization of $H$ in $f(Q,T)$ gravity }
\author{Viraj Kalsariya}
\email{virajrk6@gmail.com}
\affiliation{PD Patel Institute of Applied Sciences, Charusat University, Anand 388421, Gujarat, India}

\author{Shibesh Kumar Jas Pacif}
\email{shibesh.math@gmail.com}
\affiliation{Centre for Cosmology and Science Popularization (CCSP), SGT University, Delhi-NCR, Gurugram 122505, Haryana, India}

\begin{abstract}
Our objective in this paper is to study the late-time behavior of the universe in a model resulting from a parametrization of the Hubble parameter ($H$) in $f(Q,T)$ gravity. We have considered the flat Friedmann-Lemaitre-Robertson-Walker (FLRW) as the background metric and discussed the model in $f(Q,T)$ gravity, where $Q$ and $T$ are non-metricity and the trace of the energy-momentum tensor respectively. The complicated field equations are solved in a model-independent way by using a simple parametrization of $H$. Some geometrical parameters and physical parameters for the obtained model are calculated and their cosmic evolution are described through some graphical representation. The physical dynamics of the model are discussed in some detail. Finally, we found the model's validity by checking the energy conditions, kinematic behavior, and the speed of the sound for the obtained models from the parametrization of $H$. The interesting results of the models are compelling to the present scenario of late-time cosmic acceleration.
\end{abstract}

\maketitle

\preprint{APS/123-QED}


\affiliation{International Centre for Cosmology (ICC) \\
Charotar University of Science and Technology \\
Anand, Gujarat, India}

\affiliation{The Thanu Padmanabhan Centre for Cosmology and Science Popularization (CCSP) \\
Shree Guru Govind Singh Tricentenary (SGT) University, Gurugram \\
NCR Delhi, India}




\section{\label{sec:level1}Introduction}

Einstein's general theory of relativity (GR) is by far the most successful theory which could describe the universe in a more subtle way. It changed the perspective of the physicist about the understanding of the universe. However, the biggest problem with GR is the late-time cosmic acceleration. The observation of the supernovae of type Ia (SNIa) \cite{SupernovaCosmologyProject:1998vns, 
SupernovaSearchTeam:1998fmf} along with baryonic acoustic oscillations (BAO) \cite{BOSS:2016wmc}, acoustic peaks of cosmic microwave background radiation (CMBR) \cite{Planck:2018vyg}, and direct measurements of Hubble parameters (OHD) \cite{Magana:2017nfs} support to the late-time acceleration phenomena. This gives rise to the growing negative pressure term that is produced by the mysterious force called dark energy. The most favorable approach in GR to explain dark energy is the cosmological constant ($\Lambda$). The $\Lambda$ Cold Dark Matter ($\Lambda$CDM) model is the most basic model having the equation of state (EoS) $\omega=-1$ to explain the accelerated expansion and dust matter ($\omega= 0$) for the dark matter evolution. Although it shows auspicious results with the match of the observational data, it contains two precarious problems at the local scale which do not let us consider it as the final paradigm to describe the dynamics of the universe \cite{Peebles:2002gy,Nojiri:2010wj}. One such issue is a fine-tuning problem which refers to the dissimilarity between the predicted and the observed vacuum energy density \cite{Weinberg:1988cp}. Besides, another fundamental issue with $\Lambda CDM$ is the coincidence problem. It addresses the fact that the density of the fluid causing the cosmic acceleration and pressure-less matter are very similar at the current time even though both evolved differently during the evolution of the universe \cite{Arkani-Hamed:2000ifx,Kunz:2012aw}. Furthermore, in support of DE, many different models are introduced with different equations of states. The most famous models are Chaplygin gases \cite{Hernandez-Almada:2018osh}, Axion \cite{Peccei:2006as, Berenji:2016jji}, scalar fields \cite{Matos:2000ss, Matos:2000ng, Matos:2000ki, Urena-Lopez:2000ewq}, etc.

Recently, there has been new theories have come to light. The new generation of the dark energy model comes from the modification in Einstein-Hilbert's action. Hence, it is the logical explanation of the GR. This phenomenological method is referred to as Modified Gravity. Modified gravity theories are capable enough to comprehend the problems with the DE model by reconstructing gravitational field theory which demystifies the late-time cosmic acceleration. The first modified gravity model $f(R)$ gravity by Nojiri and Odinstov \cite{Nojiri:2003ft} was introduced by substituting the arbitrary function Ricci scalar (R) in Einstein-Hilbert action. Soon after, Harkov et al. \cite{Harko:2011kv} carried out the research further with the inclusion of a trace of energy-momentum tensor (T) in Einstein-Hilbert action. The model $f(R, T)$ is extensively studied by researchers as it has a contribution to both the matter and geometrical content.

The newly proposed modified gravity model by Yixin Xu, et al. \cite{Xu:2019sbp} is $f(Q,T)$ gravity, which is the extension of symmetric teleparallel gravity $f(Q)$ by Jimenez et al. \cite{BeltranJimenez:2017tkd}. The arbitrary functions $Q$ and $T$ are non-metricity and a trace of the energy-momentum tensor respectively. This new theory has considerable interest to study the late universe and is studied by many researchers in recent times in different contexts such as finding observational constraints \cite{Arora:2020tuk}, FLRW cosmology in $f(Q,T)$ gravity \cite{Godani:2021mld}, energy conditions \cite{Arora:2020iva, Arora:2021jik, Arora:2020met}, Cosmological implications of its Weyl-type gravity \cite{Xu:2020yeg, Gadbail:2021kgd,Gadbail:2021fjf}.

In the current paper, we have considered a parameterized model given by Pacif, et al. from \cite{Pacif:2020hai} within the classical gravity and extend the analysis in $f(Q,T)$ gravity. The purpose of the study is to check the validity of the energy conditions for both models in the $f(Q,T)$ gravity and study the evolution of the universe from the physical parameters. Additionally, we have done some cosmographic tests with the use of a model-independent approach in order to differentiate a better-performing model when compared with cosmological data \cite{Visser:1997tq}. In this approach, we take the Taylor series expansion to the scale factor (a) and find the redshift-dependent geometrical parameters like Hubble parameter (H), deceleration parameter (q), jerk parameter (j), snap parameter (s), lerk parameter (l). We also investigated some distance measurements through kinematic tests i.e., lookback time ($t_L$), proper distance ($d_p$), luminosity distance ($d_l$), and angular diameter distance ($d_A$) \cite{Arbab:1998uw}. Besides, the squared speed of the sound is also studied to find the stability of the models.

The paper is organized in the following manner. In the first section, we have introduced the current understanding of modern cosmology and possible scenarios which could describe the late-time cosmic acceleration of the universe and the paper overview. In the second section, we introduced $f(Q,T)$ gravity and associated field equations. The modified Friedmann equations are also obtained in the same section. In section III, we have discussed the parametrization and the model obtained in a model-independent way. The physical parameters of the model are explained in section IV. We have obtained the energy conditions for both the models and its plot with the variation in model parameters in the fifth section. The sixth section of the paper is associated with the interpretation of geometrical parameters, wherein we have discussed the evolution of cosmographic parameters. In the seventh section, we extended our analysis by investigating some kinematic behavior of the universe for both models. In section eight, we derive the equations for the squared speed of sound ($c_s^{\;2}$) for models M1 and M2 in order to find the stability of the models. We then conclude our analysis by concluding remarks and results in the last section.

\section{Cosmological equation in $f(Q,T)$ gravity}

From the modification in Einstein-Hilbert action of general relativity by introducing a function of two scalar invariants, $Q$ and $T$, which are constructed from the non-metricity and the trace of the energy-momentum tensor. The action in $f(Q,T)$ is given by \cite{Xu:2019sbp}.

\begin{equation}
S=\int \left[\frac{1}{16\pi}{f(Q, T)}+\mathcal{L}_M\right]\sqrt{-g}\,d^{4}x  
\label{action}
\end{equation}
where, $g\equiv det(g_{\mu \nu})$, and $\mathcal{L}_M$ is Lagrangian density.\\

In Riemannian geometry, the metric tensor is always symmetric. However, for our research, we take the $f(Q,T)$ gravity, in which, we can take the non-symmetric part of the metric tensor called non-metricity. Which can be defined as 
\begin{equation}
Q\equiv -g^{\mu \nu }(L_{\beta \mu }^{\alpha }L_{\nu \alpha }^{\beta}-L_{\beta \alpha }^{\alpha }L_{\mu \nu }^{\beta })
\end{equation}
where 
\begin{equation}
    L_{\beta \gamma }^{\alpha }\equiv -\frac{1}{2}g^{\alpha \lambda }({\nabla}_{\gamma}g_{\beta \lambda}+{\nabla }_{\beta}g_{\lambda \gamma}-{\nabla }_{\lambda}g_{\beta \gamma})
\end{equation}
\newline
The trace of nonmetricity tensor is, 
\begin{equation}
Q_{\alpha }\equiv Q_{\alpha } \\
^{\mu } \\
_{\mu },\;{\tilde{Q}}_{\alpha }\equiv Q^{\mu } \\
_{\alpha \mu }
\end{equation}
The trace of the energy-momentum tensor and modification in the metric tensor
are respectively 
\begin{equation}
T_{\mu \nu }=-\frac{2}{\sqrt{-g}}\frac{\delta (\sqrt{-g}\mathcal{L}_{M})}{%
\delta g^{\mu \nu }}
\end{equation}%
\begin{equation}
\Theta _{\mu \nu }=g^{\alpha \beta }\frac{\delta T_{\alpha \beta }}{\delta
g^{\mu \nu }}
\end{equation}
Finding the variation of action of the field equation (\ref{action}) with respect to metric tensors.

\begin{widetext}
\begin{equation}
        8\pi T_{\mu \nu}=-\frac{2}{\sqrt{-g}}\nabla_\alpha ({f}_Q \sqrt{-g}P^{\alpha}_{\mu\nu}-\frac{1}{2}{f}g_{\mu\nu}+{f}_T (T_{\mu\nu}+\Theta_{\mu\nu})\\-{f}_Q (P_{\mu\alpha\beta} Q_{\nu}^{\alpha\beta}-2Q^{\alpha\beta_{\mu}}P_{\alpha\beta\nu}))
\end{equation}
Where super-momentum 
\begin{equation}
        P^{\alpha}_{\mu\nu}\equiv\frac{1}{4} \left[-Q^{\alpha}_{\mu\nu}+2Q_{(\mu} \,^{\alpha} \,_{\nu)} +Q^{\alpha} g_{\mu \nu}- \tilde{Q}^{\alpha} g_{\mu \nu}- \delta^{\alpha}\,_{(\mu}Q_{\nu)} \right]=-\frac{1}{2} L^{\alpha}_{\mu\nu}+\frac{1}{4}\left(Q^{\alpha}-\tilde{Q}^{\alpha} \right)g_{\mu\nu}-\frac{1}{4}\delta^{\alpha}\,_{(\mu}Q_{\nu )}
\end{equation}
\end{widetext}
Taking the FLRW metric as follows, 
\begin{equation}
ds^2=-N(t)^2dt^2+a(t)^2(dx^2+dy^2+dz^2),
\end{equation}
where N(t) is the lapse function and a(t) is the scale factor. Hence, $Q=6H^2/N^2$. We assume the value of N(t) = 1, for a standard case.
Hence, $Q={6H^2}$.\\
To find the generalized Friedmann equations, assuming the matter content as the perfect fluid with the energy-momentum tensor $T_{\nu }^{\mu}=diag(-\rho,p,p,p)$. The tensor $\Theta _{\nu }^{\mu }$ becomes, 
\begin{equation}
\Theta _{\nu }^{\mu }=\delta _{\nu }^{\mu }p-2T_{\nu }^{\mu }=diag(2\rho+p,-p,-p,-p)
\end{equation}
For simplicity, taking $F\equiv {f}_{Q}= dF/dt$ and $8\pi \tilde{G}\equiv {f}_{T}=dF/dt$ the Friedmann equations we derived as
follows, 
\begin{equation}
8\pi \rho =\frac{{f}}{2}-6FH^{2}-\frac{2\tilde{G}}{1+\tilde{G}}(\dot{F}H+F\dot{H})  \label{f1}
\end{equation}
\begin{equation}
8\pi p=-\frac{{f}}{2}+6FH^{2}+2(\dot{F}H+F\dot{H})  \label{f2}
\end{equation}
From equations (\ref{f1}) and (\ref{f2}), modified Einstein's field equations are derived 
\begin{equation}
3H^{2}=8\pi \rho _{eff}=\frac{{f}}{4F}-\frac{4\pi }{F}[(1+\bar{G})\rho +\bar{G}p]  \label{efe1}
\end{equation}
\begin{multline}
{2\dot{H}+3H^{2}}=-8\pi p_{eff}=\frac{{f}}{4F}-\frac{2\dot{F}H}{F} \\
+\frac{4\pi }{F}\left[ (1+\bar{G})\rho +(2+\bar{G})p\right]  \label{efe2}
\end{multline}
From equations (\ref{efe1}) and (\ref{efe2}) and the derivative of equation (\ref{efe1}) we derive the continuity equation 
\begin{equation}
    \dot{\rho}_{eff}+3H\left( {\rho }_{eff}+{p}_{eff}\right) =0
\end{equation}

Although there are several forms of the function $f(Q,T)$ is considered in the literature \cite{Xu:2020yeg}, we here only confined to the linear and additive form of $f(Q,T)$ function \cite{Xu:2019sbp, Arora:2020iva} in the form
\begin{equation}
f(Q,T)=\mu Q+\nu T  \label{model}
\end{equation}
where, $\mu $ and $\nu $ are the non zero model constants. Hence the first derivatives ${f}_{Q}=\mu $ and $8\pi \tilde{G}={f}_{T}=\nu $.

Solving the modified Friedmann equations and applying the barotropic equation of states $p=\omega \rho$ we can find the equation of state parameter as follows
\begin{equation}
    \omega = \frac{3H^2 (8\pi+\nu) + \Dot{H} (16\pi +3\nu)}{\nu\Dot{H}-3H^2 (8\pi + \nu)}
    \label{OmegaRough}
\end{equation}

Hence the energy density equation turns out to be 
\begin{equation}
    \rho = \frac{-3H^2 \mu (8\pi+\nu)+\mu\nu\Dot{H}}{2(4\pi +\nu)(8\pi +\nu)}
    \label{Rho}
\end{equation}
To find the value of $\Dot{H}$ we use the relation $a_0 / a=1+z$ we can define a new relation between z and t. 
\begin{equation}
    \frac{d}{dt}=\frac{dz}{dt}\frac{d}{dz}=-(1+z)H(z) \frac{d}{dz}
\end{equation} 
normalizing the equation by taking the value of the scale factor as $a_0=a(0)=1$. Hence we can write the derivation Hubble parameter with respect to time in terms of red-shift as, 
\begin{equation}
    \Dot{H} = -(1+z)H(z) \frac{dH}{dz}
    \label{tzdot}
\end{equation}

\section{The Model}

In literature, the model-independent way approach is well motivated specifically in the study of dark energy models. In this approach, a model of the universe can be reconstructed and the cosmic evolution can be described mathematically, without violating the background physics. A wide variety of various parametrization schemes have been summarized and the motivation for the cosmological parametrization is discussed in \cite{Pacif:2020hai}. A simple parametrization of the Hubble parameter is considered that discusses a few known models is discussed in \cite{Pacif:2016ptv}. Following the same motivation, we here consider the parametrization of H that describes two models showing some intriguing features of the late universe. The functional form of H is considered in the form, 
\begin{equation}
H(t)=\frac{k_{2}t^{m}}{(t^{n}+k_{1})^{p}}
\end{equation}
here $k_{1},\,k_{2},\,n,\,m,\,p$ are the model parameters. Different values of $n,\,m,$and $\,p$ give rise to different models such as power law cosmology, $\Lambda CDM$ model, and models showing bouncing features. In this research, we have used the well-described two models in which the constants are $m=-1,\,p=1,\,$and $\,n=1$ and $m=-1,\,p=1,$\thinspace and $\,n=2$, out of twelve models described in \cite{Pacif:2016ptv}. It has been used
because these two models show the possibility of describing the phenomena of cosmological phase transition. We hence name them the models, M1 and M2, which has the functional forms of $H(t)$ as,\\
M1 is 
\begin{equation}
H(t)=\frac{k_{2}}{t(k_{1}-t)}  \label{M1}
\end{equation}%
M2 is 
\begin{equation}
H(t)=\frac{k_{2}}{t(k_{1}-t^{2})}  \label{M2}
\end{equation}
As we are more concerned with the late universe, it would be better, if we express the models in terms of redshift $z$. Here, for model M1, we have, 
\begin{equation}
t(z)=k_{1}\left( 1+(\beta (1+z))^{\frac{k_{1}}{k_{2}}}\right) ^{-1}
\label{M1z}
\end{equation}
and for model M2, 
\begin{equation}
t(z)=\sqrt{k_{1}}\left( 1+(\beta (1+z))^{\frac{2k_{1}}{k_{2}}}\right) ^{-\frac{1}{2}}  \label{M2z}
\end{equation}
Now, the Hubble parameter in terms of redshift $z$ for M1 is, 
\begin{equation}
\small
H(z)=H_{0}(1+\beta ^{\alpha })^{-2}(1+z)^{-\alpha }\left[ 1+(\beta(1+z))^{\alpha }\right] ^{2}  \label{M1Hz}
\end{equation}
and for M2 is, 
\begin{equation}
\small
H(z)=H_{0}(1+\beta ^{2\alpha })^{\frac{-3}{2}}(1+z)^{-2\alpha }\left[1+(\beta (1+z))^{2\alpha }\right] ^{\frac{3}{2}}  \label{M2Hz}
\end{equation}
Where in equations (\ref{M1Hz}) and (\ref{M2Hz}), we reduced one variable by taking ${k_{1}}/{k_{2}}=\alpha $ and $\beta $ is an integrating constant. Now, we have only two model parameters $\alpha $ and $\beta $ that describe the dynamics of the model.
\begin{figure}[htbp]
    \centering
    \includegraphics[width=0.83\linewidth]{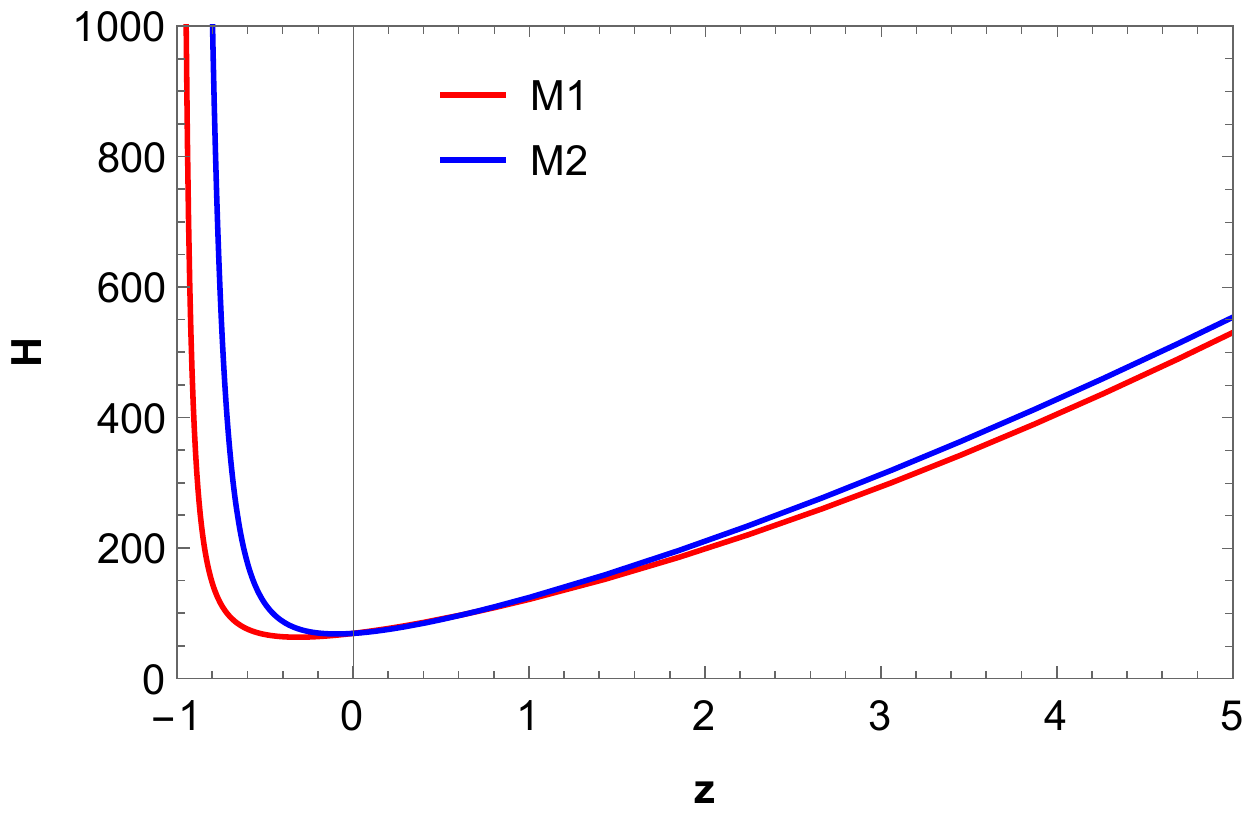}
    \caption{The Hubble parameter in terms of red-shift}
    \label{H}
\end{figure}\\
 The suitable values of these model parameters can be approximated using any observational dataset. Here, we use the constrained values of model parameters $\alpha $ and $\beta $ as found in \cite{Pacif:2020hai} using $H(z)+SN+BAO$ datasets. The best fit values found are $\alpha =1.59173$ and $\beta =1.45678$ for model M1 and $\alpha =1.42829$ and $\beta =1.40637$ for model M2. We also use the present value of the Hubble parameter as $H_0 = 69 km/s/Mpc$.

The Hubble parameter is a crucial quantity that characterized the speed of the expanding homogeneous and isotropic universe. In the figure, FIG. \ref{H}, it is certain that the universe was slowing down from the big bang ($z=\infty$) until a phase when the speed of expansion started rising and in the far future ($z=-1$) both the models are predicting the universe will continue expanding with an increasing rate of velocity. Using these best-fit values of $\alpha $ and $\beta $, we plot the various cosmological parameters of our models and see the late-time behavior of the universe in model M1 and M2 in $f(Q,T)$ gravity. In the following section, we shall discuss the Energy conditions for our models in the considered $f(Q,T)$ gravity.

\section{Evolution of Physical Parameters}

In this section, we shall find the physical parameters for the models in the $f(Q,T)$ gravity. From that, we can check the model's energy condition. These energy conditions are fundamental and valuable to characterize the matter in the universe.

The EoS parameter is associated with energy density $\rho$ and pressure density $p$. It is useful to classify the expansion of the universe. When $\omega=1$, it represents stiff fluid. $\omega=0$ is an indication of a matter-dominated universe and radiation-dominated when $\omega=1/3$. On the other hand, $-1<\omega<0$, represents the quintessence phase however, the EoS parameter below -1 ($\omega<-1$) shows the phantom era. Also, the cosmological constant could be seen at $\omega=-1$.\\
We find the EoS parameter for models M1 and M2 from Eq. (\ref{OmegaRough}). Which can be written as\\
for M1
\begin{multline}
  \omega = -  \Big[\big((1+z)\beta \big)^\alpha\Big(8\pi(3-2\alpha)+3\nu(1-\alpha)\Big)+
  \\
    \Big(8\pi(3+2\alpha)+3\nu(1+\alpha)\Big) \Big]/ \Big[\Big(24\pi+3\nu-\alpha\nu\Big) 
  \\
  +\big((1+z)\beta\big)^\alpha \Big(24\pi+3\nu+\alpha\nu\Big) \Big]
  \label{M1 omega}
\end{multline}
for M2
\begin{multline}
    \omega = - \Big[\big((1+z)\beta \big)^{2\alpha} \Big(8\pi(3-2\alpha)+3\nu(1-\alpha)\Big)+
    \\
    \Big(8\pi(3+4\alpha)+ 3\nu(1+\alpha)\Big)\Big]/ \Big[\Big(24\pi+3\nu-2\alpha\nu\Big) 
    \\    +\big((1+z)\beta\big)^{2\alpha}\Big(24\pi+3\nu+3\alpha\nu\Big)\Big]
    \label{M2 omega}
\end{multline}

\begin{figure}[htbp]
    \centering
    \includegraphics[width=0.85\linewidth]{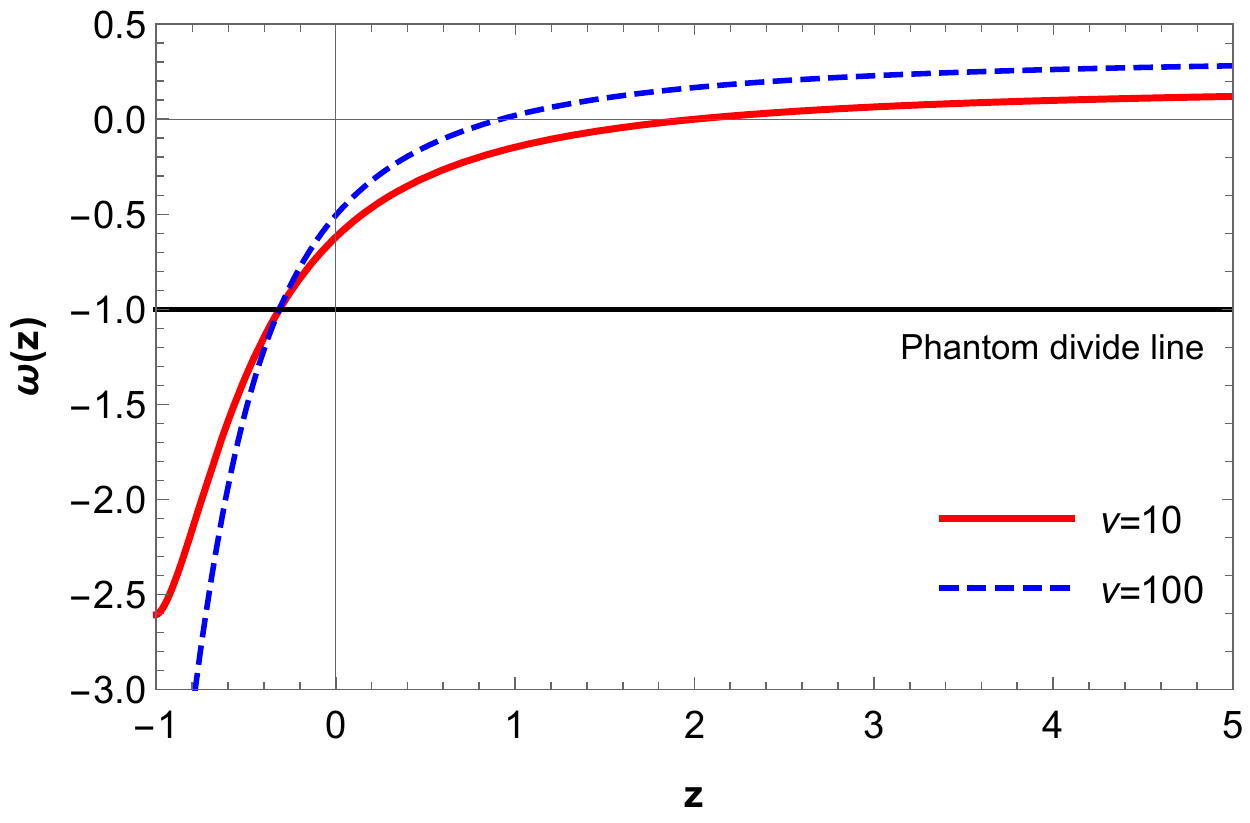}
    \caption{Equation of state parameter for model M1}
    \label{fig.M1omega}
    \includegraphics[width=0.85\linewidth]{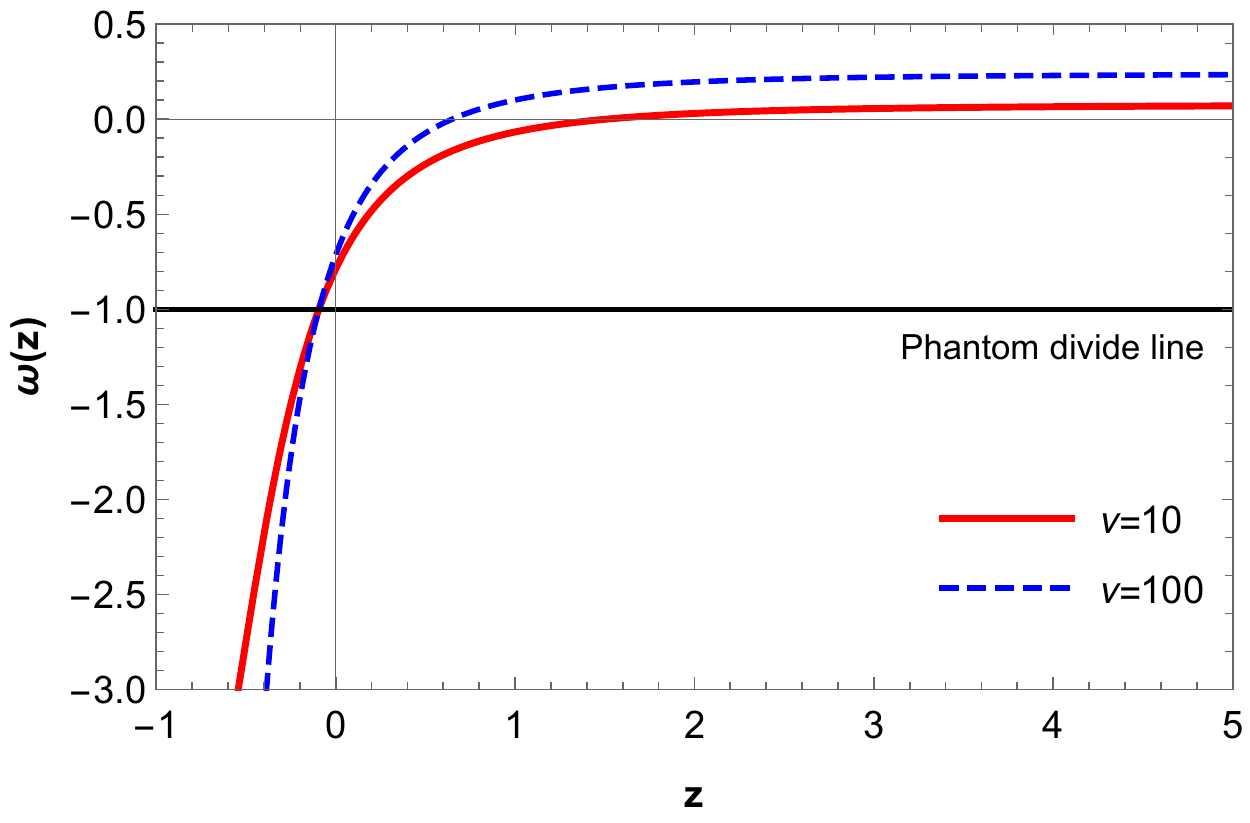}
    \caption{Equation of state parameter for model M2}
    \label{fig.M2omega}
\end{figure}
With the constrained values of the model parameters, we plotted the EoS parameter for both models. We found that in order to explain the standard evolution of the EoS parameter as predicted by the Planck2015 and Planck2018 results, we need to fix the values of the considered $f(Q,T)$ form model parameters $\nu$ as $\nu>-2.6$ and $\nu>2.81$ for models M1 and M2 respectively. So, for our analysis, we have taken the values $\nu=10$ and $\nu=100$. With these criteria, we have shown the evolution of the EoS parameter $\omega$ in figure, FIG. \ref{fig.M1omega} and FIG. \ref{fig.M2omega}.\\
In Fig. (\ref{fig.M1omega}) for model M1, the value of EoS is $\omega_0=-0.6203$ for $\nu=10$ and $\omega_0=-0.5058$ for $\nu=100$ at $z=0$. On the other hand, for model M2 in Fig. (\ref{fig.M2omega}) at $z=0$ when $\nu=10$, $\omega_0 = -0.7878$ and when $\nu=100$ the value of EoS is $-0.7149$.
Now solving the energy density equations from Eq. (\ref{Rho})using the parametrization functions for the models M1 and M2,\\
for M1
\begin{equation}
\small
    \begin{split}
    \rho(z)=\Big[-H_0^2 \mu\Big(1+((1+z)\beta)^\alpha\Big)^{3} \Big((24\pi +3\nu -\alpha) 
    \\
    +((1+z)\beta)^\alpha (24\pi +3\nu +\alpha) \Big)\Big] / 
    \\
    \Big[2(1+z)^{2\alpha}(1+\beta^\alpha)^4 (4\pi+\nu)(8\pi +\nu)\Big]
    \label{EDM1}
    \end{split}
\end{equation}
for M2
\begin{equation}
\small
    \begin{split}
    \rho(z)=\Big[-H_0^2 \mu \Big(1+((1+z)\beta)^{2\alpha}\Big)^{2}\Big((24\pi +3\nu -2\alpha\nu) 
    \\
    +((1+z)\beta)^{2\alpha} (24\pi +3\nu +2\alpha\nu) \Big)\Big] /
    \\
    \Big[2(1+z)^{4\alpha}(1+\beta^{2\alpha})^3 (4\pi+\nu)(8\pi +\nu)\Big]
    \label{EDM2}     
    \end{split}
\end{equation}

\begin{figure}[htbp]
    \centering
    \includegraphics[width=0.85\linewidth]{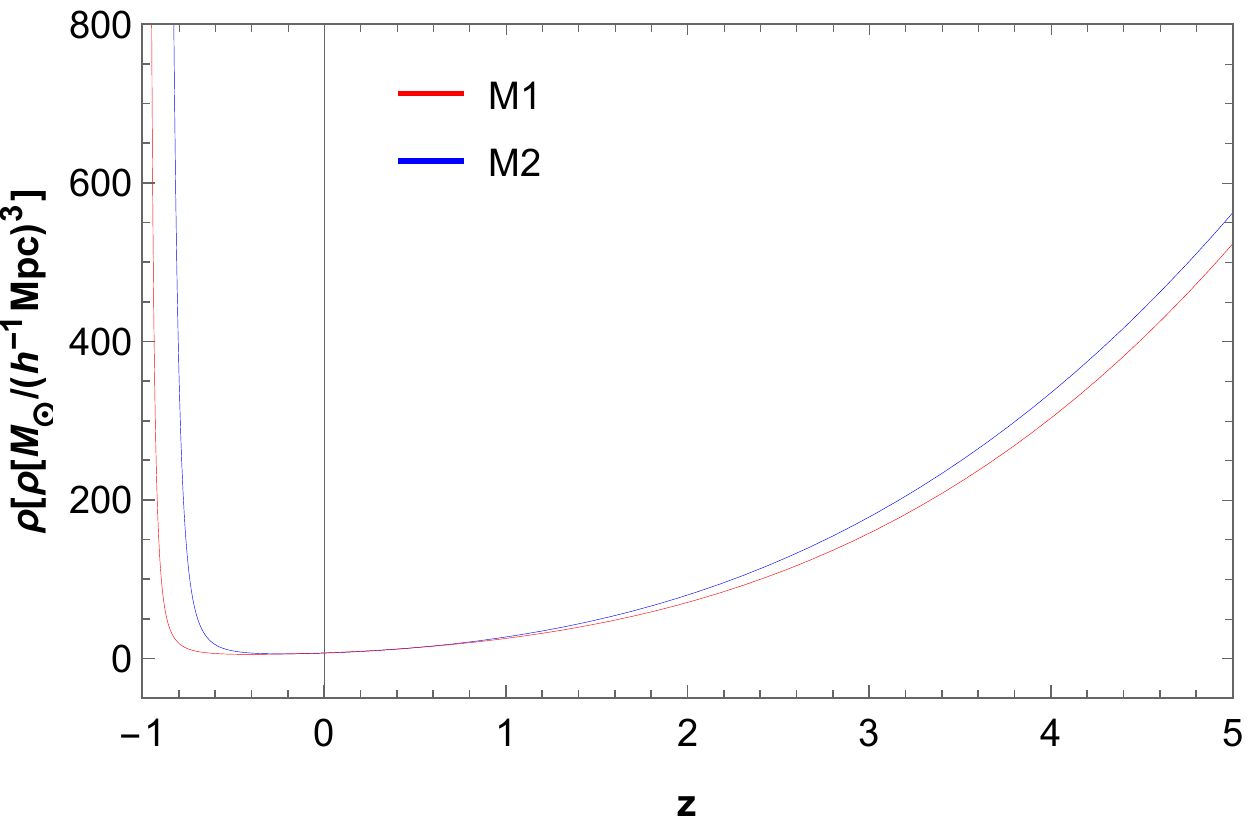}
    \caption{Energy density versus redshift for models M1 and M2}
    \label{fig.rho}
\end{figure}
Fig. (\ref{fig.rho}) portray the behavior of the energy density in terms of the redshift plot for models M1 and M2 while considering model parameters $\mu=-0.1$, $\nu=100$, and $H_0=69$. The energy density is found to be positive for the full range of redshift.\\

\section{Energy Conditions}

Continuing our analysis by obtaining the energy conditions from the Raychaudhury equation from \cite{Raychaudhuri:1953yv}. The energy conditions for the models' validity can be stated below, 
\begin{enumerate}
\item Strong energy condition (SEC): Gravity is always attractive. 
\begin{align}
SEC:\rho+3p \geq 0
\end{align}

\item Dominant energy condition (DEC): The cannot travel faster than the
speed of light. 
\begin{align}
DEC: \rho \geq |p|
\end{align}

\item Weak energy condition (WEC): The energy density should always be
non-negative. 
\begin{align}
WEC: \rho \geq 0 \quad and \quad \rho+p \geq 0
\end{align}

\item Null energy condition (NEC): A minimum requirement for SEC and WEC. 
\begin{align}
NEC: \rho+p \geq 0
\end{align}
\end{enumerate}

We now find the validity of the energy conditions for both models separately.

\subsection{EC for Model M1}
Defining the energy conditions for model M1 using the energy density equation in (\ref{EDM1}) and EoS parameter equation in (\ref{M1 omega}). 
\begin{widetext}
\small
\renewcommand{\arraystretch}{1.2}
\begin{multline}
    SEC: \frac{-H_0^2 \mu \Big(1+((1+z)\beta)^\alpha\Big)^{3}\Big[(24\pi +3\nu -\alpha) +((1+z)\beta)^\alpha (24\pi +3\nu +\alpha) \Big]}{2(1+z)^{2\alpha}(1+\beta^\alpha)^4 (4\pi+\nu)(8\pi +\nu)}
    \\
    +\frac{3H_0^2 \mu \Big(1+((1+z)\beta)^{\alpha}\Big)^{3} \Big[(8\pi(3+ 2\alpha)+3\nu(1+\alpha)+ \Big((1+z)\beta)^\alpha\Big)\Big(8\pi(3- 2\alpha)+3\nu(1-\alpha)\Big) \Big]}{2(1+z)^{2\alpha}(1+\beta^{\alpha})^4 (4\pi+\nu)(8\pi +\nu)} \geq 0 
\end{multline}
\begin{multline}
    DEC :\frac{-H_0^2 \mu \Big(1+((1+z)\beta)^\alpha\Big)^{3} \Big[(24\pi +3\nu -\alpha) +((1+z)\beta)^\alpha (24\pi +3\nu +\alpha) \Big]}{2(1+z)^{2\alpha}(1+\beta^\alpha)^4 (4\pi+\nu)(8\pi +\nu)}
    \\ 
    \mp \frac{H_0^2 \mu \Big(1+((1+z)\beta)^{\alpha}\Big)^{3} \Big[(8\pi(3+ 2\alpha)+3\nu(1+\alpha)+ \Big((1+z)\beta)^\alpha\Big)\Big(8\pi(3- 2\alpha)+3\nu(1-\alpha)\Big) \Big]}{2(1+z)^{2\alpha}(1+\beta^{\alpha})^4 (4\pi+\nu)(8\pi +\nu)} \geq 0 
\end{multline}

\begin{multline}
     WEC : \frac{-H_0^2 \mu \Big(1+((1+z)\beta)^\alpha\Big)^{3} \Big[(24\pi +3\nu -\alpha) +((1+z)\beta)^\alpha (24\pi +3\nu +\alpha) \Big]}{2(1+z)^{2\alpha}(1+\beta^\alpha)^4 (4\pi+\nu)(8\pi +\nu)} \geq 0 
\end{multline}

\begin{multline}
    NEC : \frac{-H_0^2 \mu \Big(1+((1+z)\beta)^\alpha\Big)^{3} \Big[(24\pi +3\nu -\alpha) +((1+z)\beta)^\alpha (24\pi +3\nu +\alpha) \Big]}{2(1+z)^{2\alpha}(1+\beta^\alpha)^4 (4\pi+\nu)(8\pi +\nu)}
    \\ 
    +\frac{H_0^2 \mu \Big(1+((1+z)\beta)^{\alpha}\Big)^{3} \Big[(8\pi(3+ 2\alpha)+3\nu(1+\alpha)+ \Big((1+z)\beta)^\alpha\Big)\Big(8\pi(3- 2\alpha)+3\nu(1-\alpha)\Big) \Big]}{2(1+z)^{2\alpha}(1+\beta^{\alpha})^4 (4\pi+\nu)(8\pi +\nu)} \geq 0 
\end{multline}
\end{widetext}
\newpage

 \begin{figure}[htbp]
    \centering
    \includegraphics[width=0.85\linewidth]{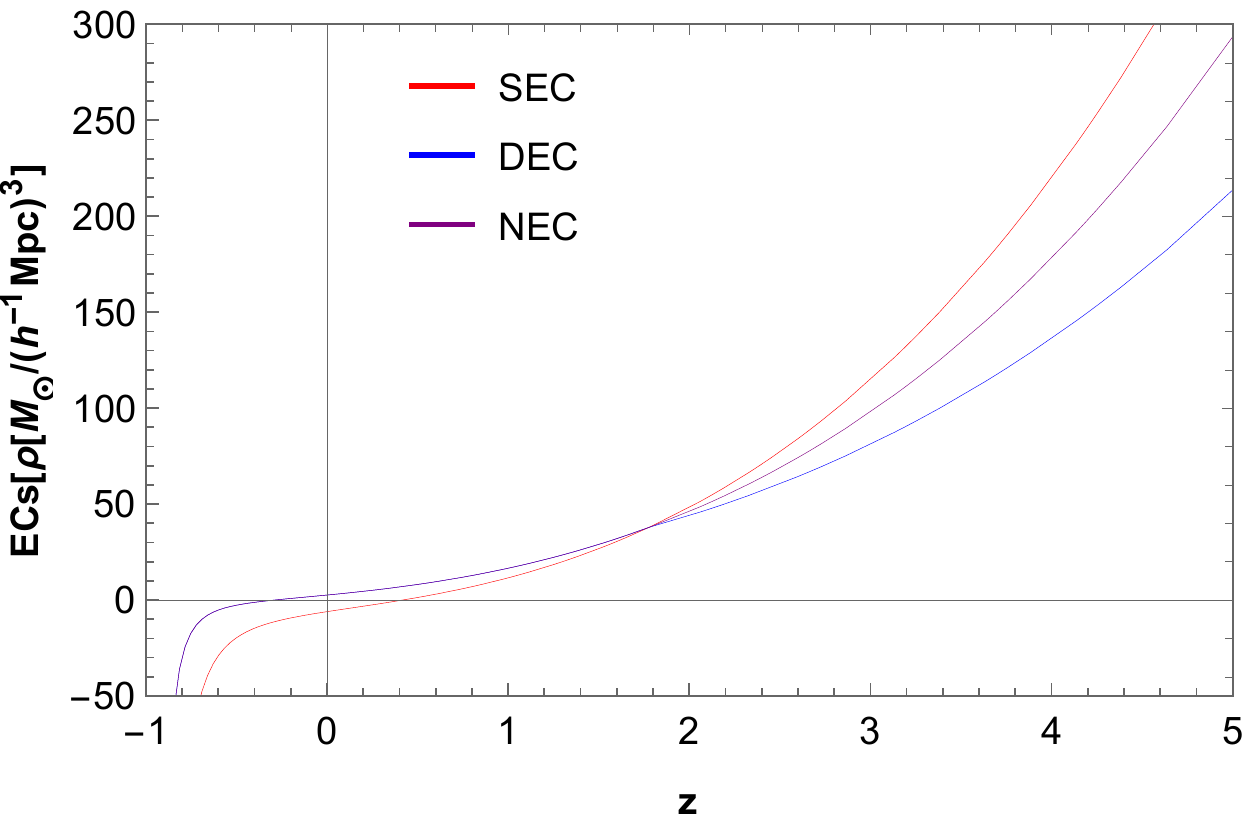}
    \caption{Energy conditions in terms of redshift for model M1}
    \label{fig.M1ECs}
\end{figure}
We have taken all the functional forms of energy conditions into account and use the appropriate value of model parameters, for instance, $\nu=10$ and $\mu=-0.1$ and the best-fit value of $\alpha= 1.5917 $ and $\beta= 1.4567$ to plot the above figure and compare the evolution in the energy conditions in terms of redshift. 
The figure (\ref{fig.M1ECs}) represents the behavior of energy conditions in the function of redshift for model M1. The plot exploits one of the most important features. From the overlapping of energy conditions, the transition of the model for which the universe's matter dominance modifies into the radiation dominance phase could be found. The transition for model M1 happens at $z=1.7812$.\\
We also observe that the SEC violates at $z=0$ which supports the most important quality of the universe, accelerated expansion of the universe. However, the DEC and NEC hold until $z<-0.2889$ showing a more intense expansion in near future.
\\We now plot the energy conditions for both variations in $\nu$ and $\mu$ and find the acceptable value of the model parameters $\nu$ and $\mu$.\\

\begin{widetext}

\begin{figure*}[htbp]
    \centering
    \vspace{1cm}
    \begin{minipage}{0.45\textwidth}
        \centering
        \includegraphics[width=1\linewidth]{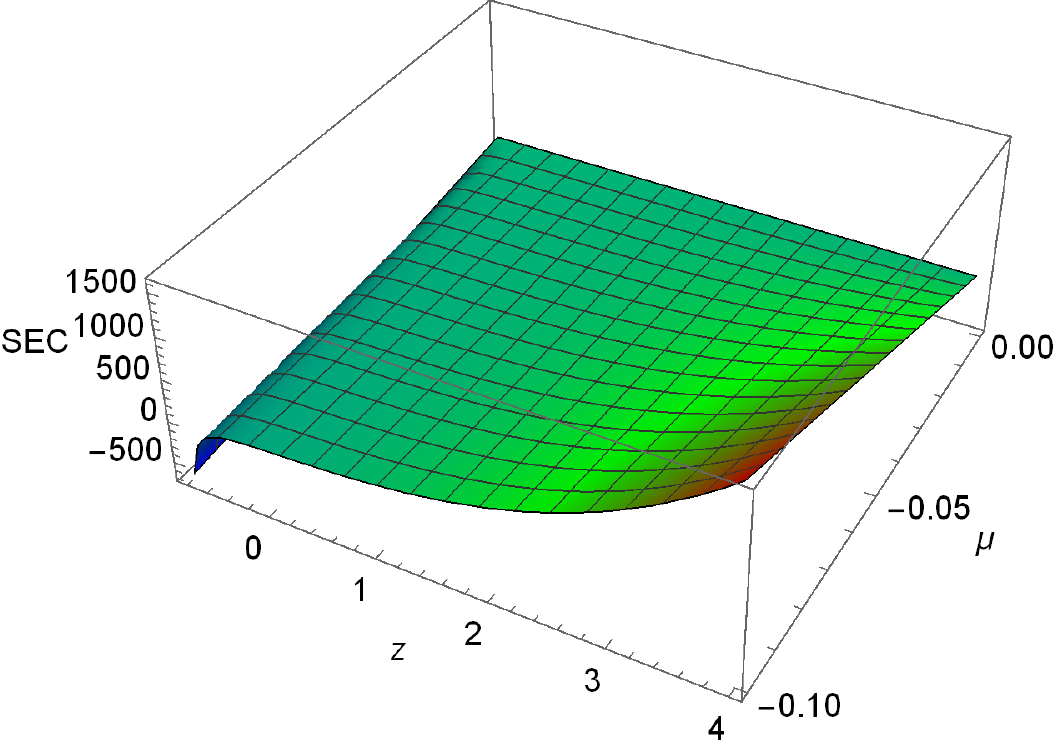}
        \caption{SEC: variation in $\mu$}
        \label{fig.M1SECmu}
    \end{minipage}\hfill
    \begin{minipage}{0.45\textwidth}
        \centering
        \includegraphics[width=0.95\linewidth]{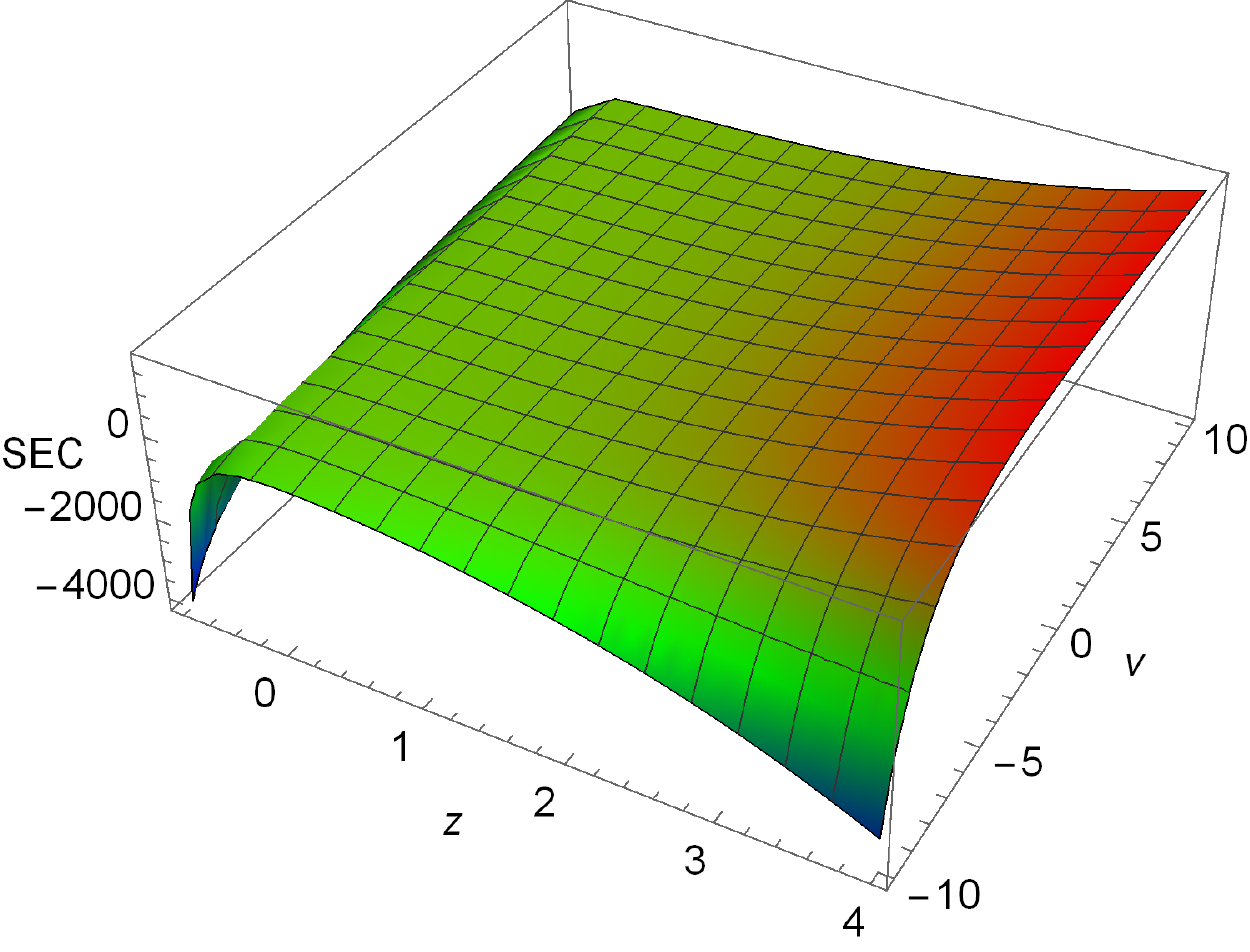}
        \caption{SEC: variation in $\nu$}
        \label{fig.M1SECnu}
    \end{minipage} 
    \vspace{1cm}
\end{figure*}
\begin{figure*}
    \begin{minipage}{0.45\textwidth}
        \centering
        \includegraphics[width=1\linewidth]{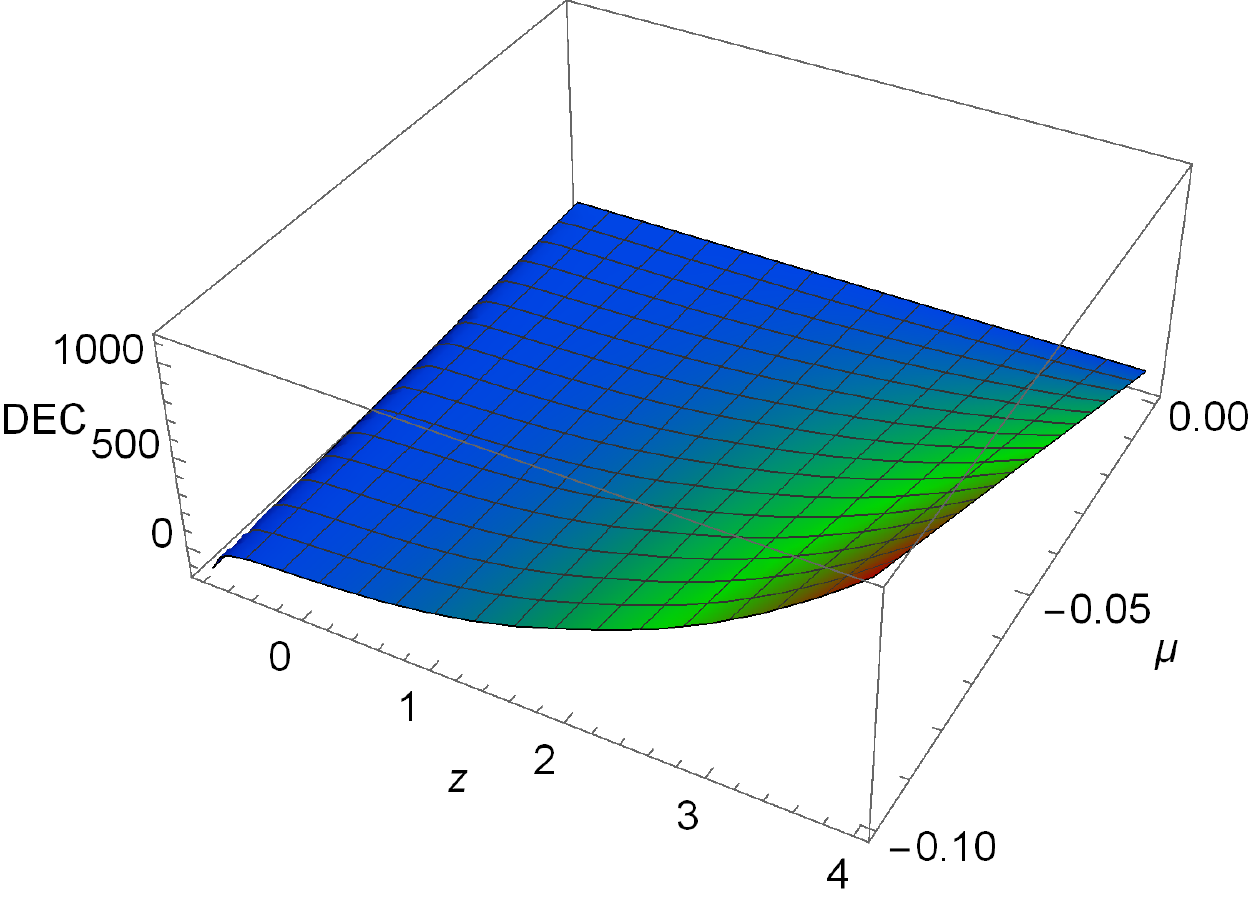}
        \caption{DEC: variation in $\mu$}
        \label{fig.M1DECmu}
    \end{minipage}\hfill
    \begin{minipage}{0.45\textwidth}
        \centering
        \includegraphics[width=1\linewidth]{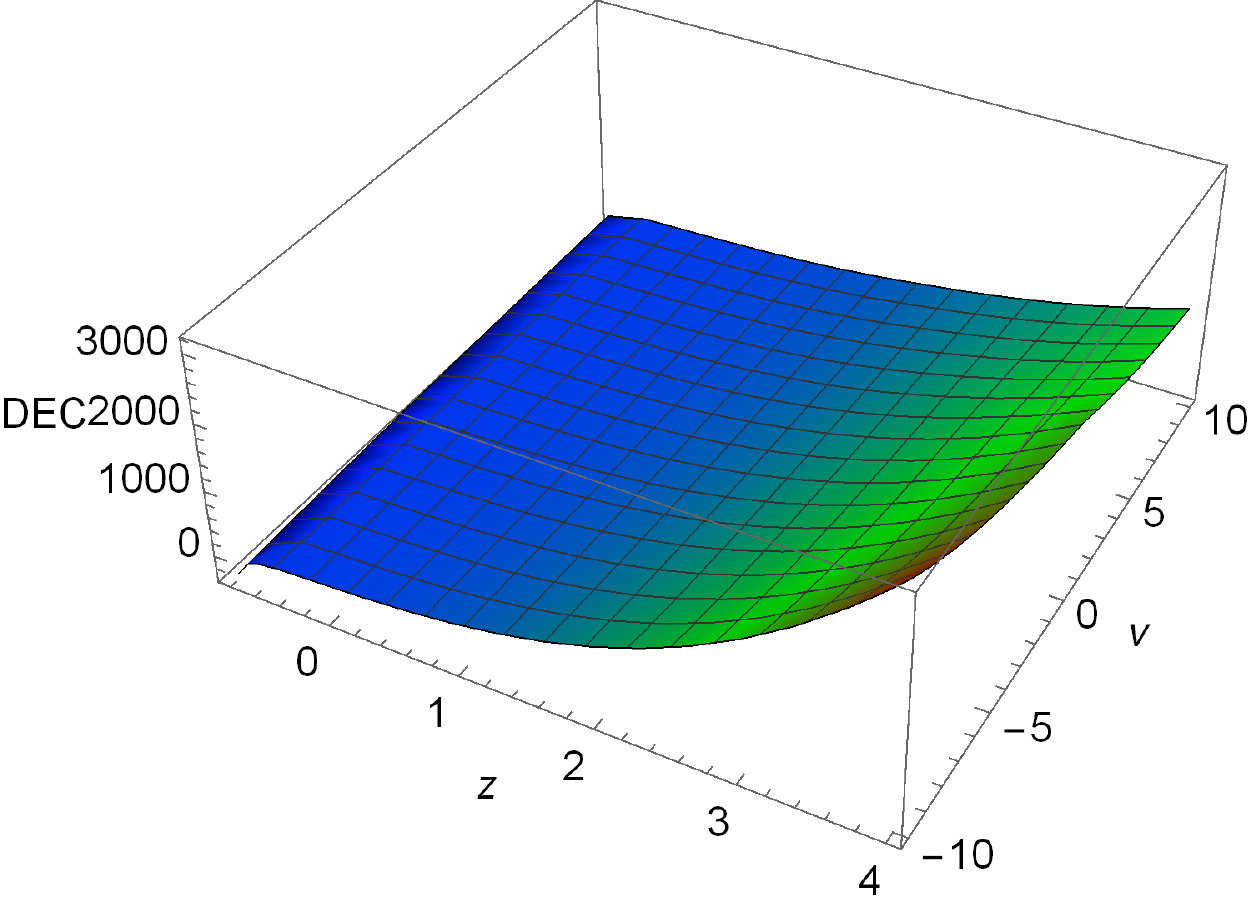}
        \caption{DEC: variation in $\nu$}
        \label{fig.M1DECnu}
    \end{minipage}
    \vspace{0.2cm}
    
    \begin{minipage}{0.45\textwidth}
        \centering
        \includegraphics[width=\linewidth]{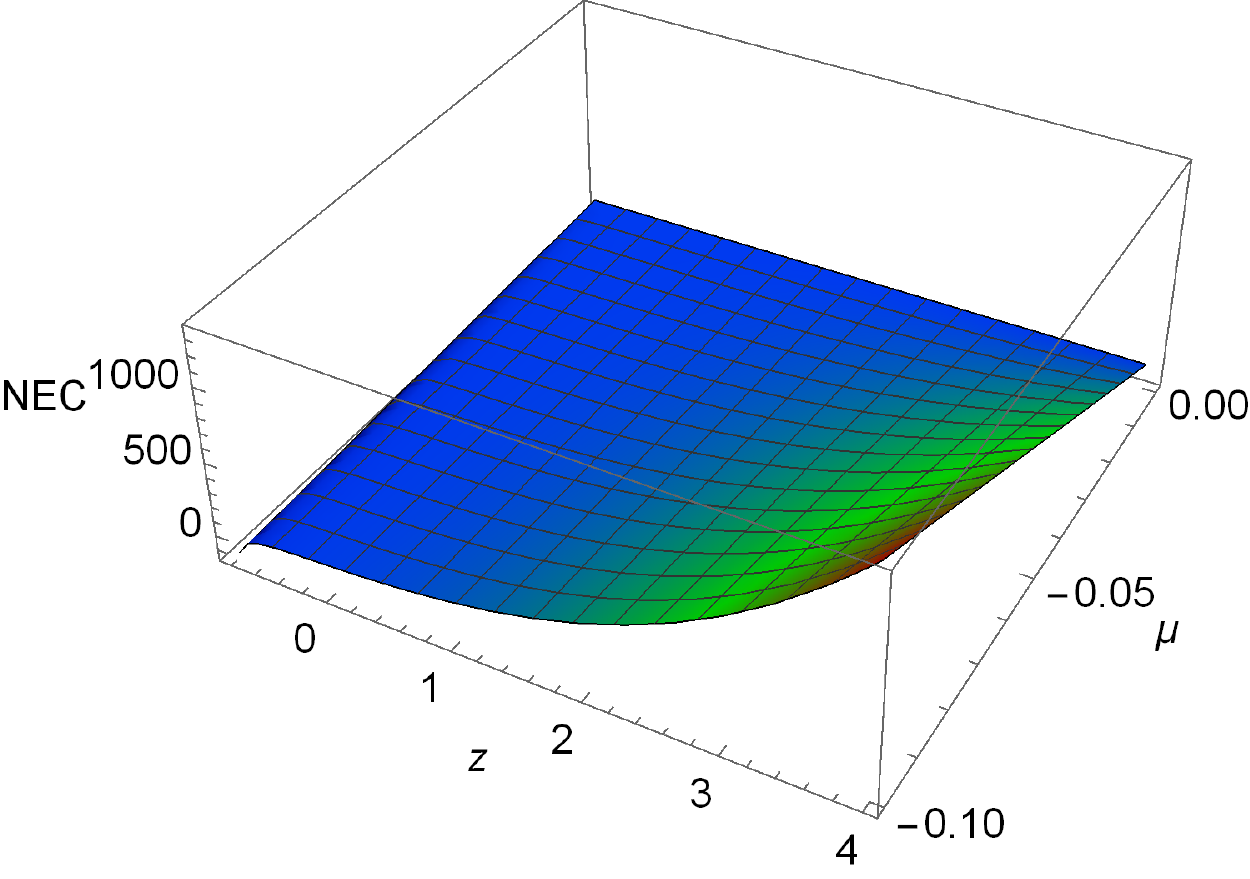}
        \caption{NEC: variation in $\mu$}
        \label{fig.M1NECmu}
    \end{minipage}\hfill
    \begin{minipage}{0.45\textwidth}
        \centering
        \includegraphics[width=\linewidth]{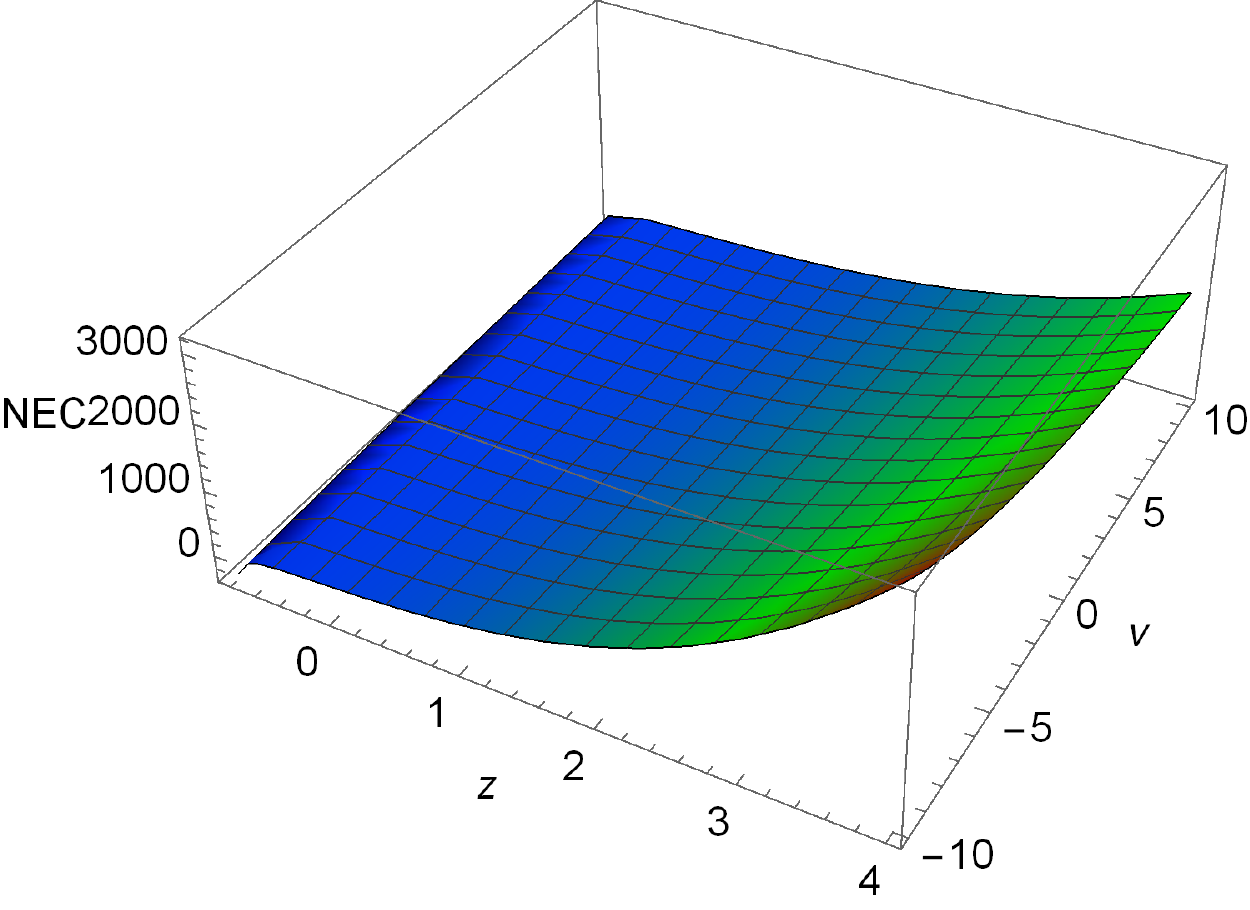}
        \caption{NEC: variation in $\nu$}
        \label{fig.M1NECnu}
    \end{minipage}
\end{figure*}
\end{widetext}

The SEC is one of the most important quantities to characterize the universe among all the energy conditions. According to the observational data of the accelerating universe, the SEC must get violated on the cosmological scale shown in \cite{Barcelo:2002bv, Moraes:2019pao}. The SEC for model M1 is shown in Figs. (\ref{fig.M1SECmu}) and (\ref{fig.M1SECnu}) with the appropriate  variation in $\mu$ and $\nu$ respectively. We observe that the feature of the SEC violation is satisfied for the range of the model parameter, $\mu<0$, and $\nu >-12.56$ at $z=0$. On the other hand, the DEC and the NEC do not violate in the given range of the model parameters which is shown in the figures (\ref{fig.M1DECmu}), (\ref{fig.M1DECnu}), (\ref{fig.M1NECmu}), and (\ref{fig.M1NECnu}). However, the DEC and the NEC do not hold for the near future.

\subsection{EC for Model M2}
Similarly, the energy conditions for model M2 can be derived using Eq. (\ref{Rho}). 

\begin{widetext}
\begin{multline}
\small
    SEC: \frac{-H_0^2 \mu \Big(1+((1+z)\beta)^{2\alpha}\Big)^{2} \Big[(24\pi +3\nu -2\alpha\nu) +((1+z)\beta)^{2\alpha} (24\pi +3\nu +2\alpha\nu) \Big]}{2(1+z)^{4\alpha}(1+\beta^{2\alpha})^3 (4\pi+\nu)(8\pi +\nu)}
    \\ 
    +\frac{3H_0^2 \mu \Big(1+((1+z)\beta)^{2\alpha}\Big)^{2} \Big[(8\pi(3+ 4\alpha)+3\nu(1+6\alpha\nu)+ \Big((1+z)\beta)^{2\alpha}\Big)\Big(8\pi(3- 2\alpha)+3\nu(1-\alpha)\Big) \Big]}{2(1+z)^{4\alpha}(1+\beta^{2\alpha})^3 (4\pi+\nu)(8\pi +\nu)} \geq 0 
\end{multline}

\begin{multline}
    DEC : \frac{-H_0^2 \mu \Big(1+((1+z)\beta)^{2\alpha}\Big)^{2}\Big[(24\pi +3\nu -2\alpha\nu) +((1+z)\beta)^{2\alpha} (24\pi +3\nu +2\alpha\nu) \Big]}{2(1+z)^{4\alpha}(1+\beta^{2\alpha})^3 (4\pi+\nu)(8\pi +\nu)}
    \\ 
    \mp \frac{H_0^2 \mu \Big(1+((1+z)\beta)^{2\alpha}\Big)^{2} \Big[(8\pi(3+ 4\alpha)+3\nu(1+6\alpha\nu)+ \Big((1+z)\beta)^{2\alpha}\Big)\Big(8\pi(3- 2\alpha)+3\nu(1-\alpha)\Big) \Big]}{2(1+z)^{4\alpha}(1+\beta^{2\alpha})^3 (4\pi+\nu)(8\pi +\nu)} \geq 0 
\end{multline}

\begin{multline}
    WEC :  \frac{-H_0^2 \mu \Big(1+((1+z)\beta)^{2\alpha}\Big)^{2} \Big[(24\pi +3\nu -2\alpha\nu) +((1+z)\beta)^{2\alpha} (24\pi +3\nu +2\alpha\nu) \Big]}{2(1+z)^{4\alpha}(1+\beta^{2\alpha})^3 (4\pi+\nu)(8\pi +\nu)} \geq 0 
\end{multline}

\begin{multline}
    NEC :\frac{-H_0^2 \mu \Big(1+((1+z)\beta)^{2\alpha}\Big)^{2}\Big[(24\pi +3\nu -2\alpha\nu) +((1+z)\beta)^{2\alpha} (24\pi +3\nu +2\alpha\nu) \Big]}{2(1+z)^{4\alpha}(1+\beta^{2\alpha})^3 (4\pi+\nu)(8\pi +\nu)}
    \\ 
    +\frac{H_0^2 \mu \Big(1+((1+z)\beta)^{2\alpha}\Big)^{2} \Big[(8\pi(3+ 4\alpha)+3\nu(1+6\alpha\nu)+ \Big((1+z)\beta)^{2\alpha}\Big)\Big(8\pi(3- 2\alpha)+3\nu(1-\alpha)\Big) \Big]}{2(1+z)^{4\alpha}(1+\beta^{2\alpha})^3 (4\pi+\nu)(8\pi +\nu)} \geq 0 
\end{multline}
\end{widetext}

\begin{figure}
    \centering
    \includegraphics[width=0.85\linewidth]{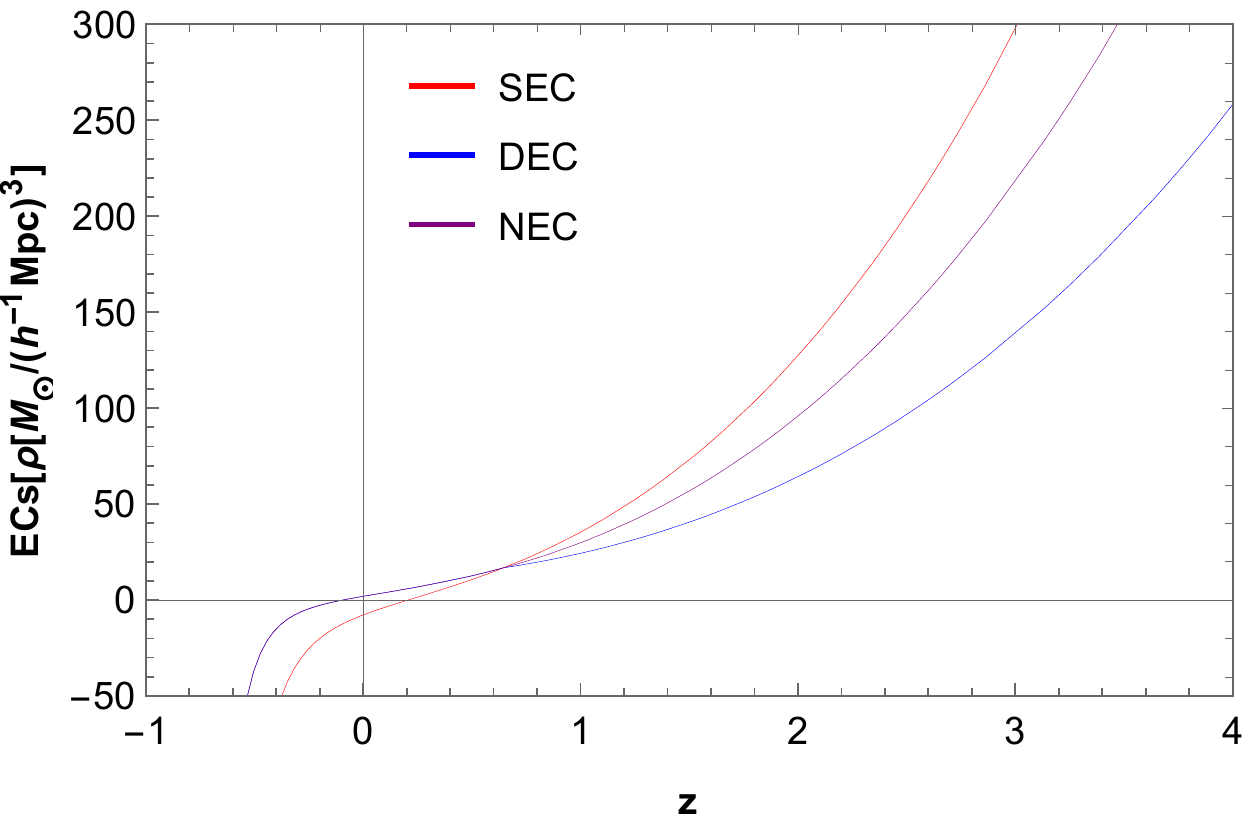}
    \caption{Energy conditions in terms of redshift for model M2}
    \label{fig.M2ECs}
\end{figure}

Fig. (\ref{fig.M2ECs}) shows the similar behavior of energy conditions of the function of redshift for model M2 as model M1 while considering the model parameters value $\nu=10$, $\mu=-0.1$, $\alpha=1.4262$, and $\beta=1.4063$. We can also observe that the SEC does not hold at $z=0$ because one can find the transition of matter dominance to radiation dominance happens at $z=0.6473$ supporting the expanding universe. However, the NEC and DEC stay positive and do hold for the current period of the universe ($z=0$). Besides, the DEC and NEC do get violated for $z<-0.0936$ suggesting the big rip universe.\\

We now plot the graphs for the variation in $\mu$ as well as $\nu$ to check for which a range of model parameters gives the solution supporting the current observed data.

\begin{widetext}

\begin{figure*}[htbp]
    \centering

    \begin{minipage}{0.45\textwidth}
        \centering
        \includegraphics[width=0.93\linewidth]{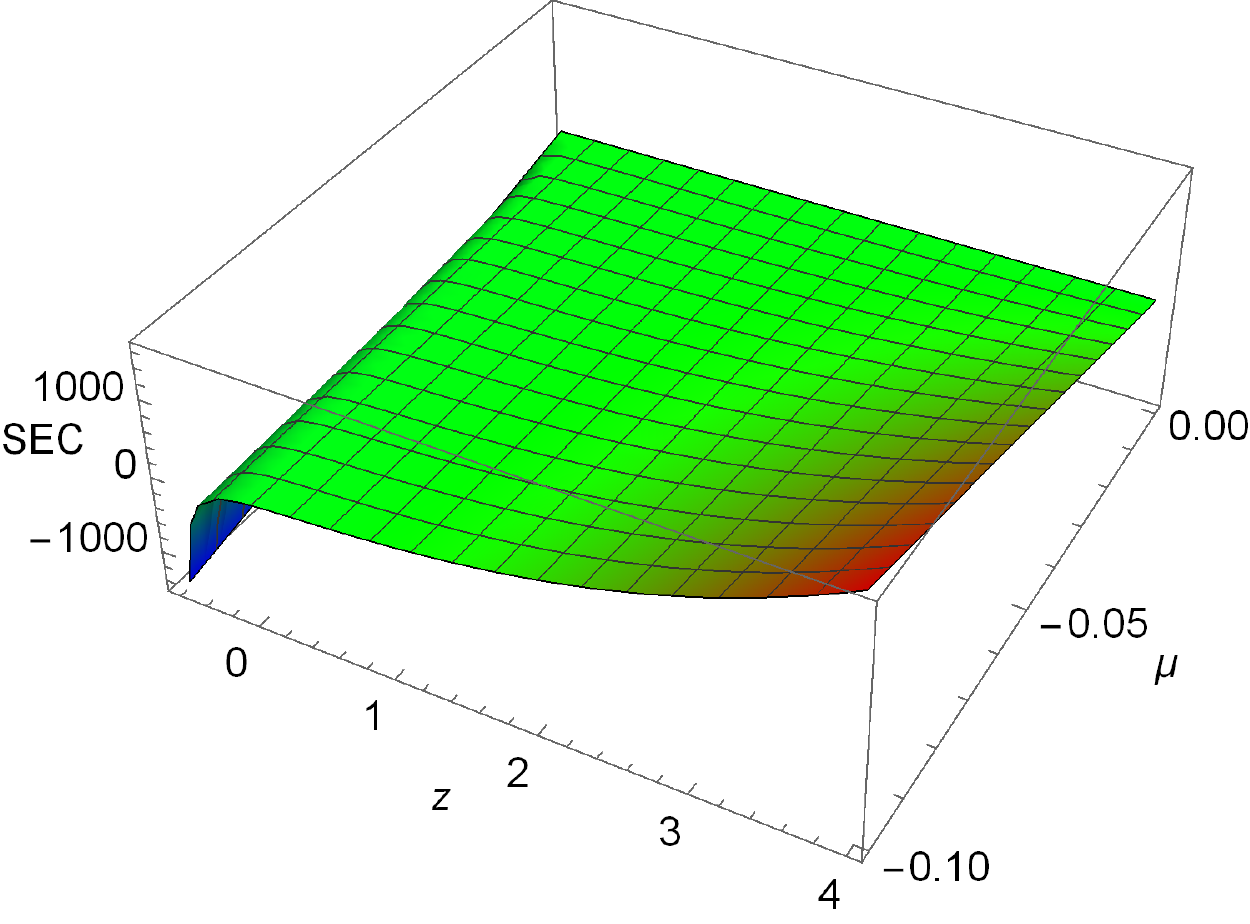}
        \caption{SEC: variation in $\mu$}
        \label{fig.M2SECmu}
    \end{minipage}\hfill
    \begin{minipage}{0.45\textwidth}
        \centering
        \includegraphics[width=0.9\linewidth]{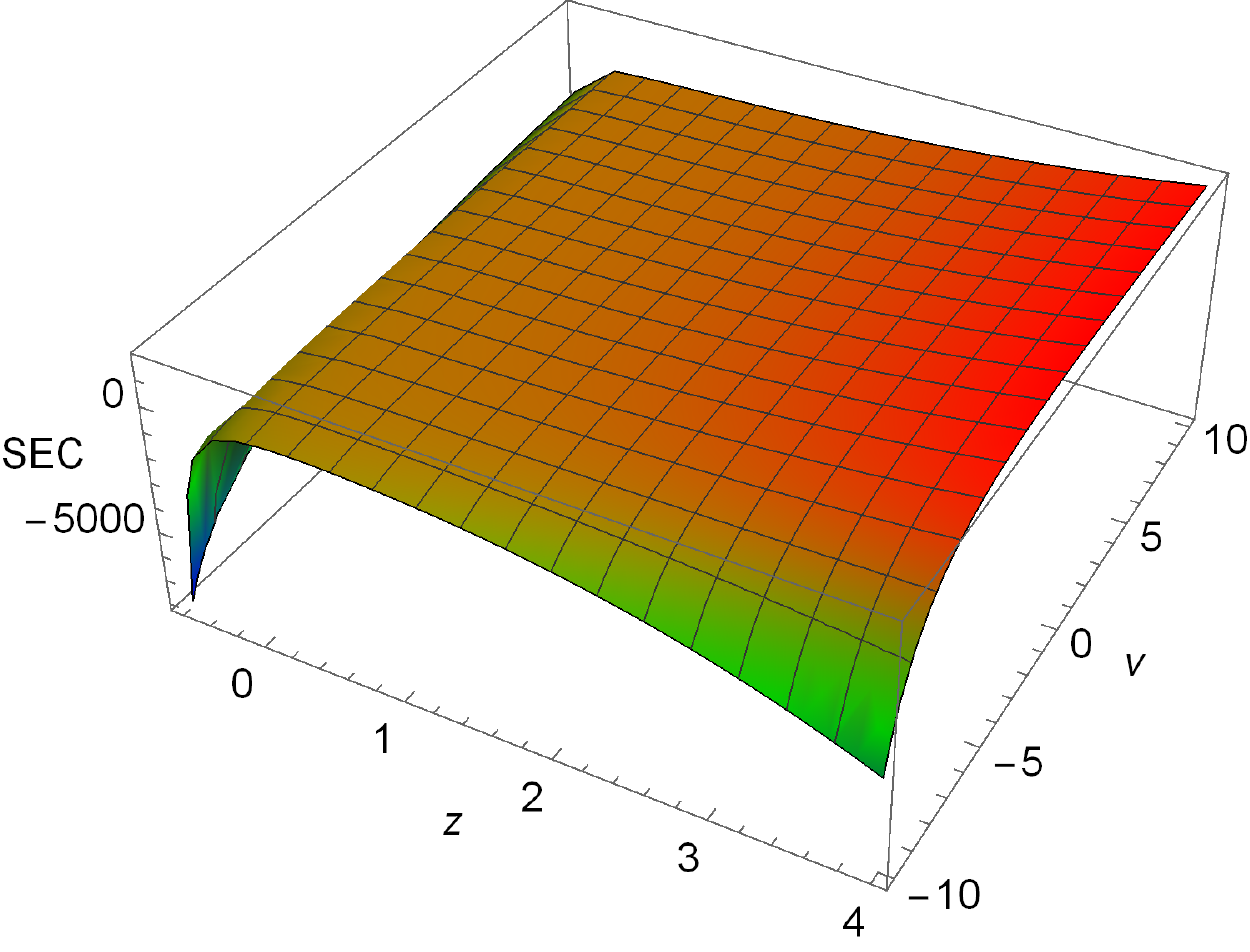}
        \caption{SEC: variation in $\nu$}
        \label{fig.M2SECnu}
    \end{minipage}
\end{figure*} 
\newpage
\begin{figure*}
    \begin{minipage}{0.45\textwidth}
        \centering
        \includegraphics[width=1\linewidth]{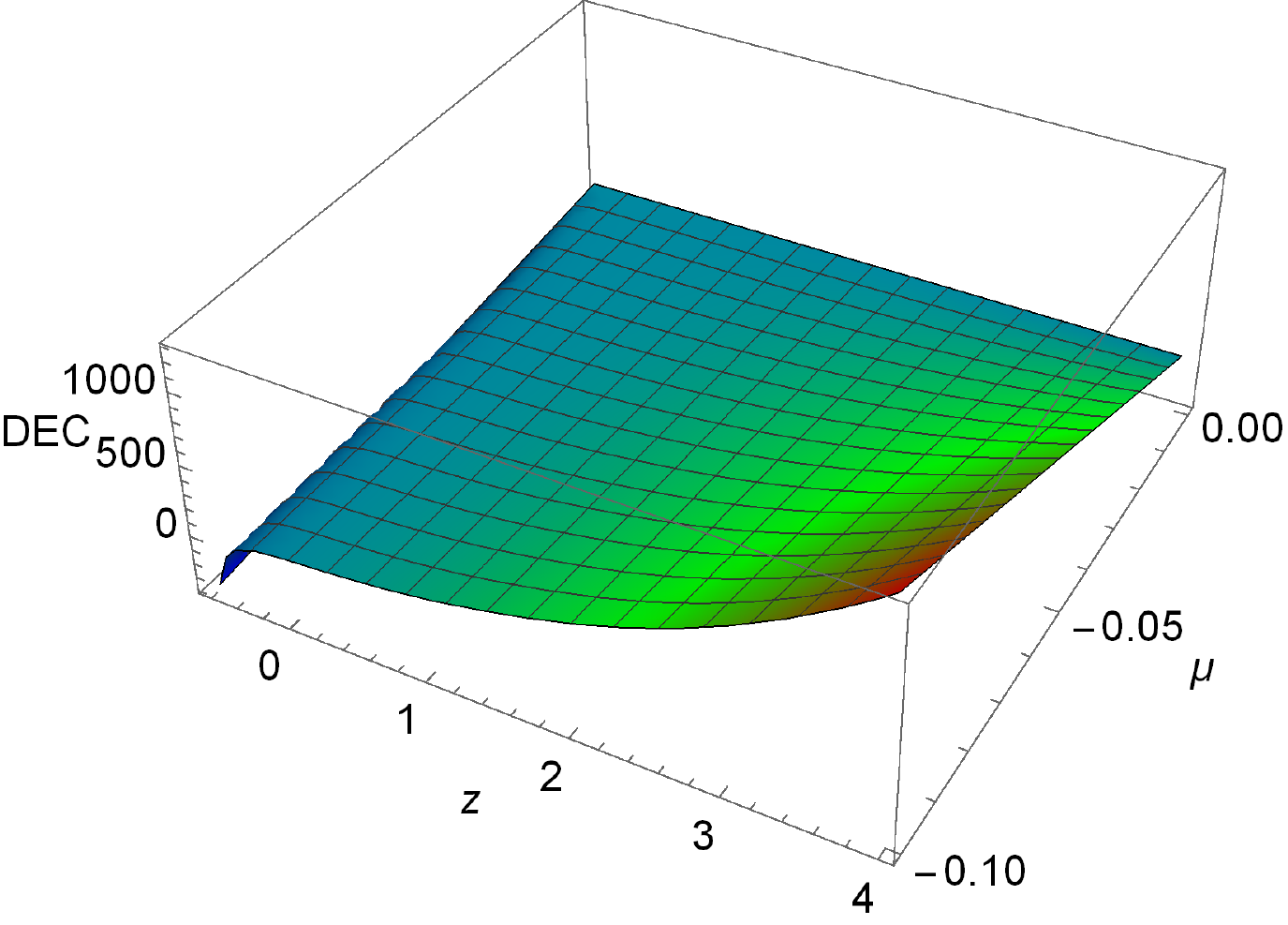}
        \caption{DEC: variation in $\mu$}
        \label{fig.M2DECmu}
    \end{minipage}\hfill
    \begin{minipage}{0.45\textwidth}
        \centering
        \includegraphics[width=1\linewidth]{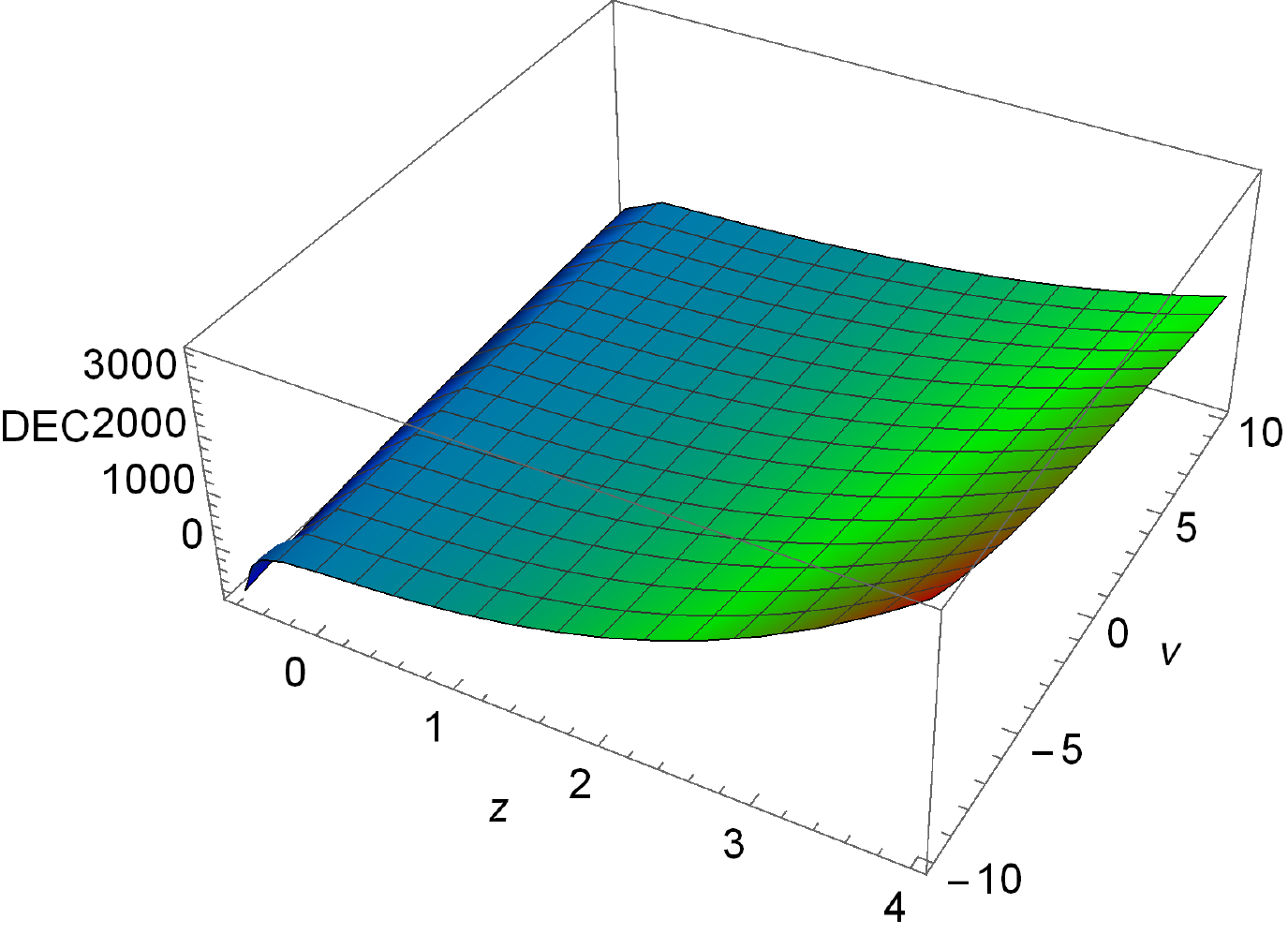}
        \caption{DEC: variation in $\nu$}
        \label{fig.M2DECnu}
    \end{minipage}

    \begin{minipage}{0.45\textwidth}
        \centering
        \includegraphics[width=\linewidth]{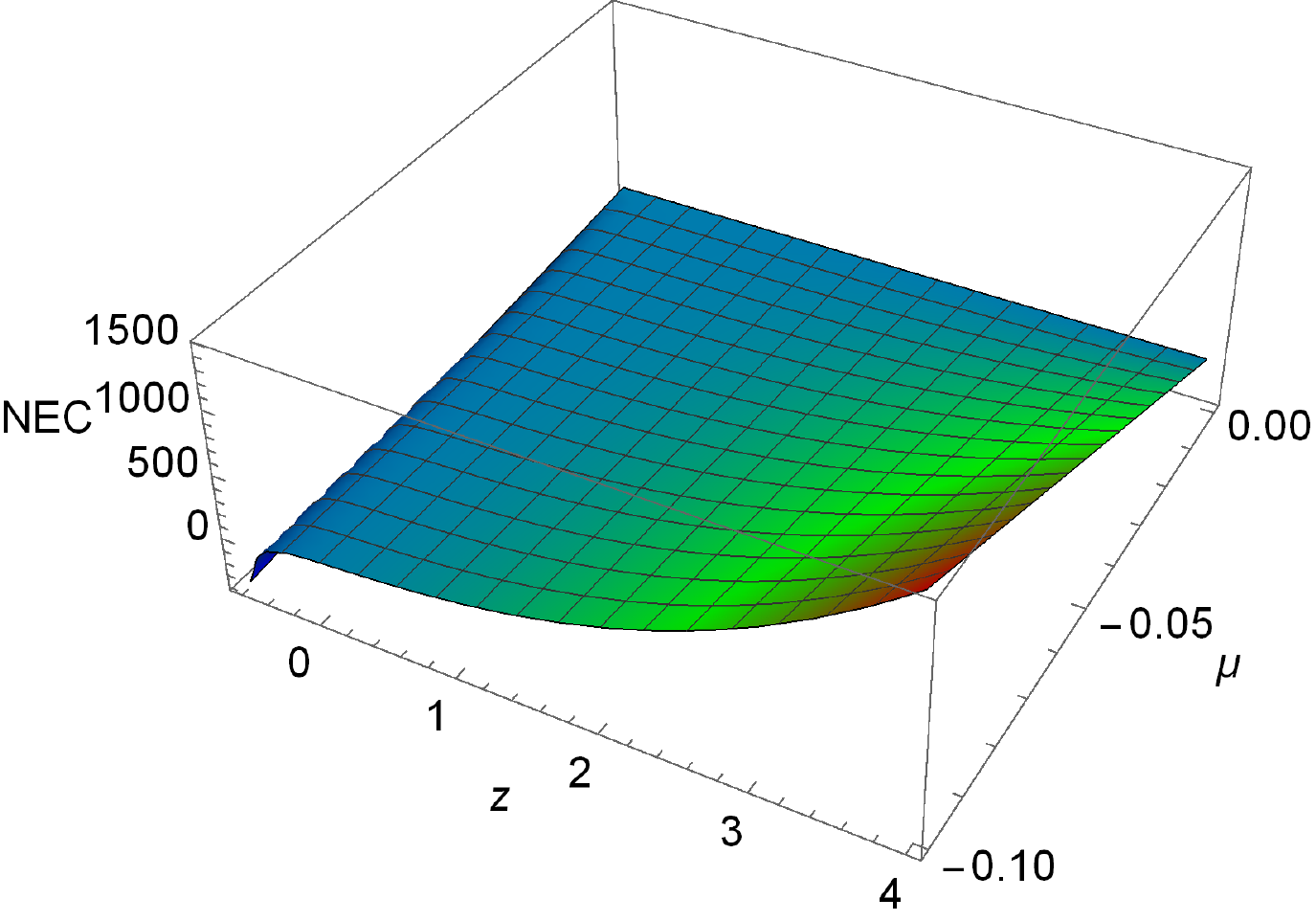}
        \caption{NEC: variation in $\mu$}
        \label{fig.M2NECmu}
    \end{minipage}\hfill
    \begin{minipage}{0.45\textwidth}
        \centering
        \includegraphics[width=\linewidth]{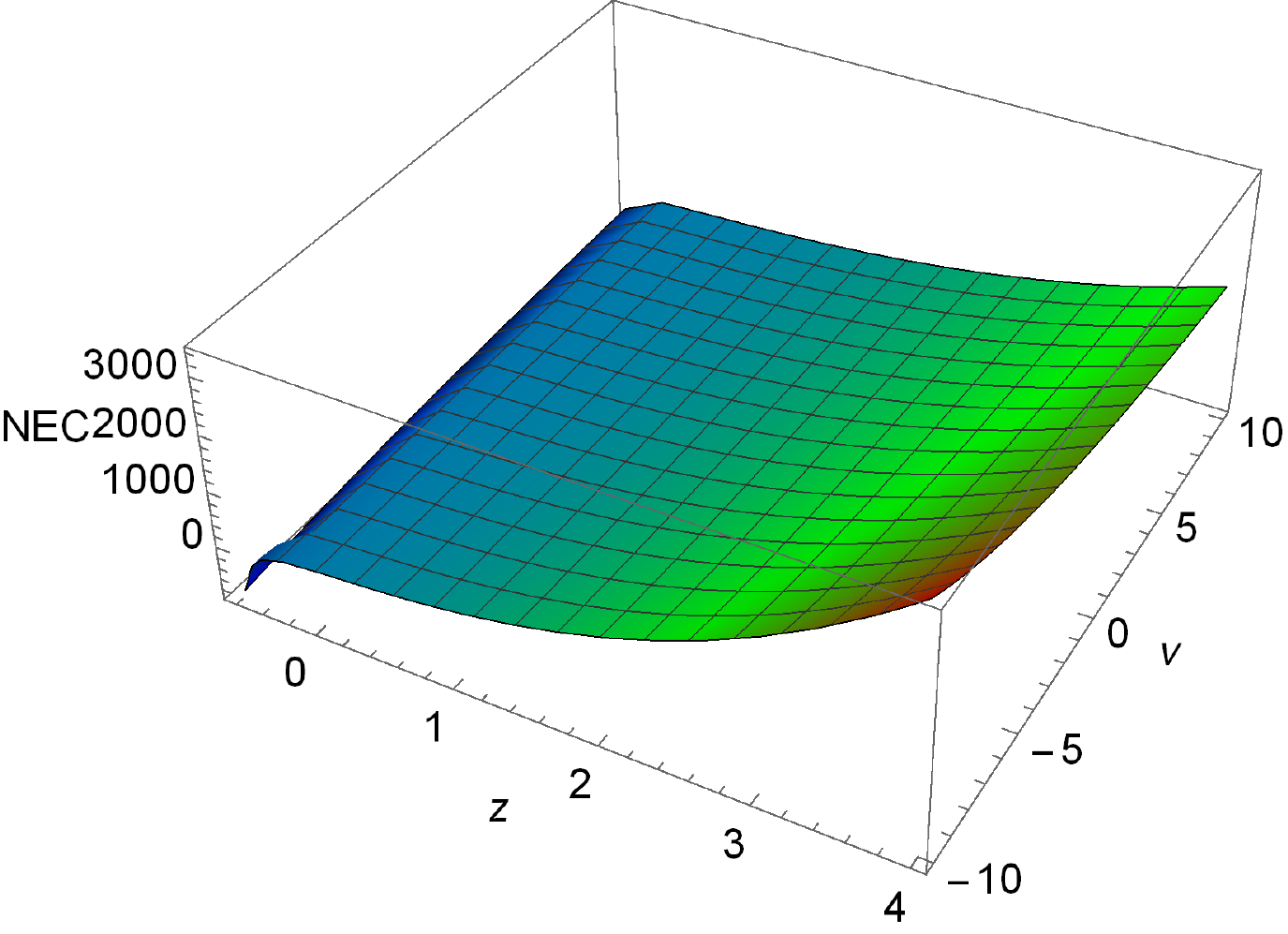}
        \caption{NEC: variation in $\nu$}
        \label{fig.M2NECnu}
    \end{minipage}
\end{figure*}
\end{widetext}

For model M2, we find a similar solution to model M1. According to the Penrose-Hawking singularity, the SEC must get violated and that supports the validity of our current solutions for models M1 and M2.  Interestingly, we found the same boundary for the model parameter. For $\nu <-12.56$, we find the singularity, on the other hand for the positive value of $\mu$, we find the mirror plot hence the condition must get satisfied in order to have a valid model. 
In the next section, we discuss the cosmic evolution of some geometrical parameters for both the models M1 and M2. 
\section{Interpretation of Geometrical Parameters}
Cosmography is used to constrain the kinematics of the universe in a model-independent way. It provides an objective means to find the correspondence of a model with observations. In the model-independent approach, we find the Taylor series expansion of the scale factor $a(t)$ which is the fundamental quantity that describes the evolution of the universe at a given time. When we do the Taylor series expansion of the scale factor, we can approximate the behavior of the scale factor to a certain degree of accuracy. 
\begin{equation}
a(t)=a_{0}\left[ 1+\sum_{n=1}^{\infty }\frac{1}{n!}\frac{d^{n}a}{dt^{n}}%
(t-t_{0})^{n}\right]  \label{tse}
\end{equation}
In the Taylor series expansion eq. (\ref{tse}) we find the cosmographic parameters of the universe. Deceleration parameter q, jerk parameter j, snap parameter s, and lerk parameter l. These cosmographic parameters are used to study the evolution and dynamics of the universe.

\begin{eqnarray*}
q=-\frac{1}{aH^2}\frac{d^2 a}{dt^2}, \quad j=\frac{1}{aH^3}\frac{d^3 a}{dt^3}, \\
s=\frac{1}{aH^4}\frac{d^4 a}{dt^4}, \quad l=\frac{1}{aH^5}\frac{d^5 a}{dt^5}
\end{eqnarray*}

In the following subsections, we are going to find the cosmographic parameters for models M1 and M2 respectively to find the evolution of the universe.\\ 

\subsection{ Evolution of Cosmographic parameters in model M1}

We use the value of our parameterized function H in terms of z from Eq. (\ref{M1Hz}) and the relation between the time and redshift from Eq. (\ref{M1z}) in the geometrical parameters to find the functional form in terms of red-shift.

\begin{equation}
    q(z)=-1+\alpha\left[1-\frac{2}{({1+(\beta(1+z))^\alpha})}\right]
\end{equation}
\begin{widetext}
\small
\begin{equation}
    j(z)=1+3\alpha\left[-1+\frac{2}{({1+(\beta(1+z))^\alpha})}\right]+\alpha^2\left[2+\frac{6}{({1+(\beta(1+z))^\alpha})^2}-\frac{6}{({1+(\beta(1+z))^\alpha})}\right]
\end{equation}
\begin{multline}
    s(z)=1+\alpha\left[-6+\frac{12}{({1+(\beta(1+z))^\alpha})}\right]+\alpha^2\left[11-\frac{36}{({1+(\beta(1+z))^\alpha})}+\frac{36}{({1+(\beta(1+z))^\alpha})^2}\right]\\+\alpha^3\left[-6+\frac{24}{({1+(\beta(1+z))^\alpha})}-\frac{36}{({1+(\beta(1+z))^\alpha})^2}+\frac{24}{({1+(\beta(1+z))^\alpha})^3}\right]    
\end{multline}
\begin{multline}
    l(z)=1+\alpha\left[10+\frac{20}{({1+(\beta(1+z))^\alpha})}\right]+\alpha^2\left[35-\frac{120}{({1+(\beta(1+z))^\alpha})}+\frac{120}{({1+(\beta(1+z))^\alpha})^2}\right]\\+\alpha^3\left[-50+\frac{220}{({1+(\beta(1+z))^\alpha})}-\frac{360}{({1+(\beta(1+z))^\alpha})^2}+\frac{240}{({1+(\beta(1+z))^\alpha})^3}\right]\\+\alpha^4\left[24-\frac{120}{({1+(\beta(1+z))^\alpha})}+\frac{240}{({1+(\beta(1+z))^\alpha})^2}-\frac{240}{({1+(\beta(1+z))^\alpha})^3}+\frac{120}{({1+(\beta(1+z))^\alpha})^4}\right]
\end{multline}

\begin{figure}[htbp]
    \centering
    \includegraphics[width=0.36\linewidth]{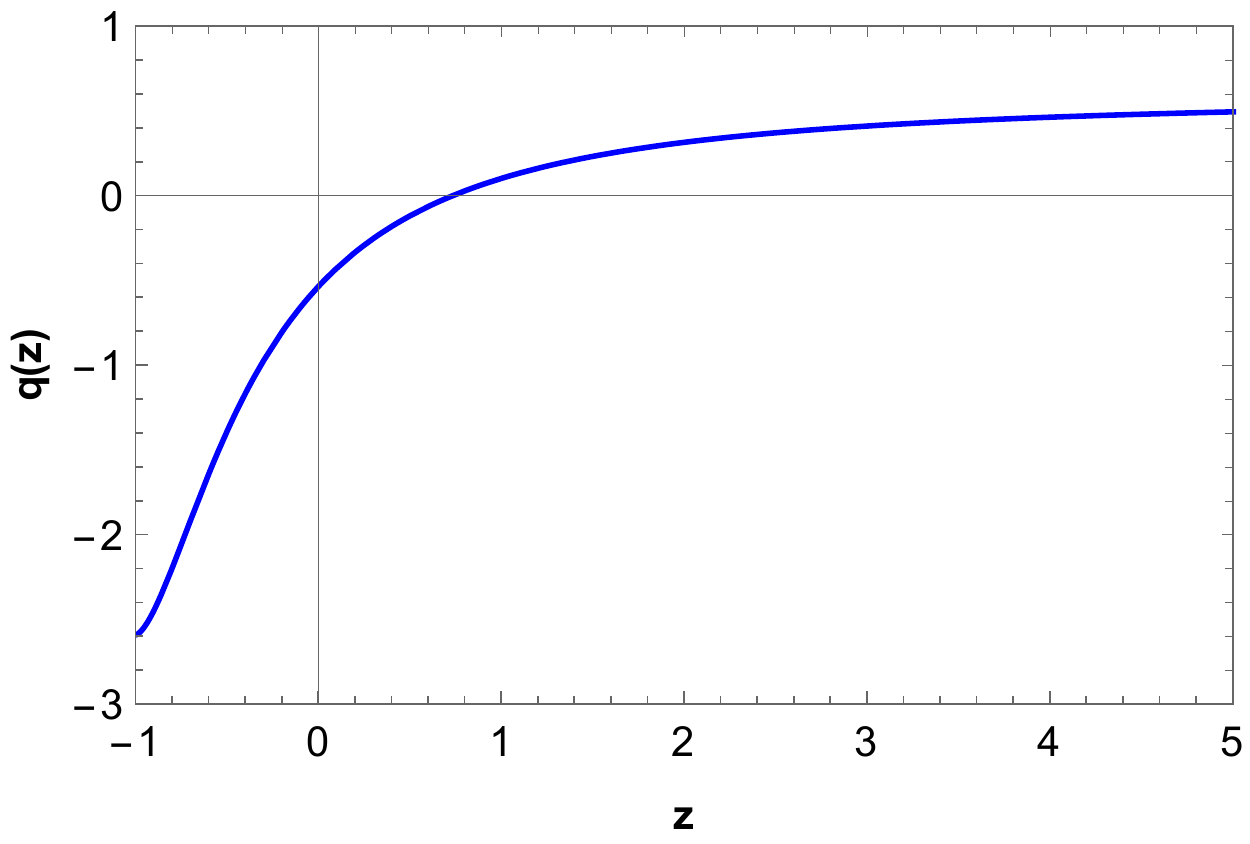}
    \includegraphics[width=0.36\linewidth]{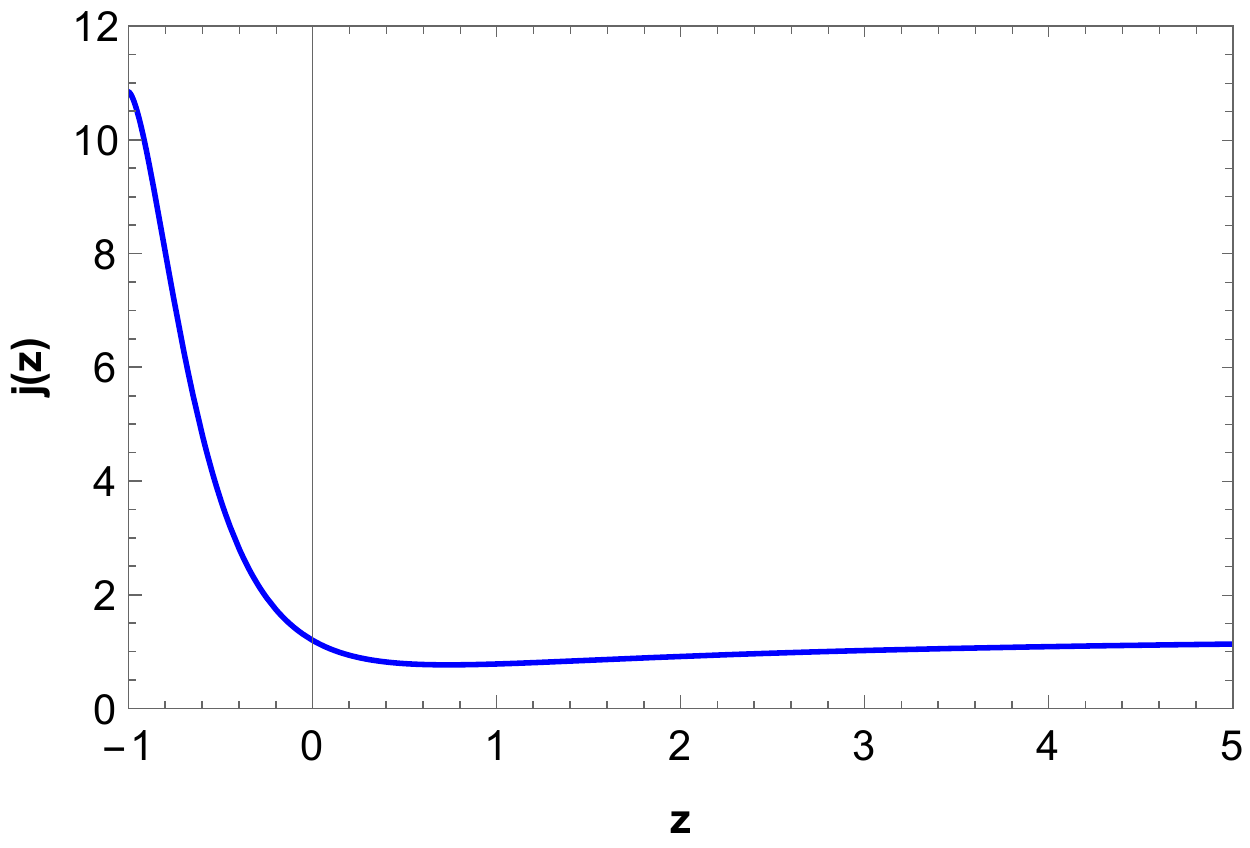}
    \includegraphics[width=0.36\linewidth]{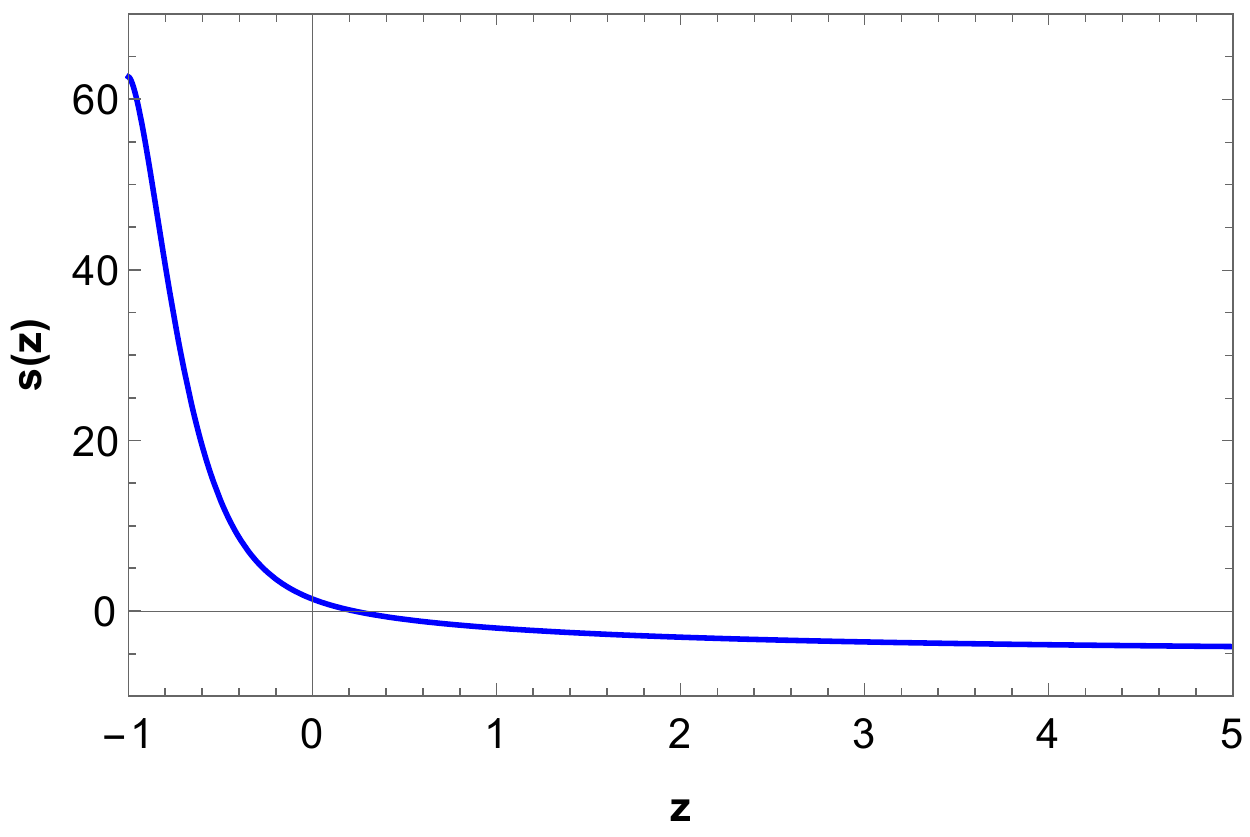}
    \includegraphics[width=0.36\linewidth]{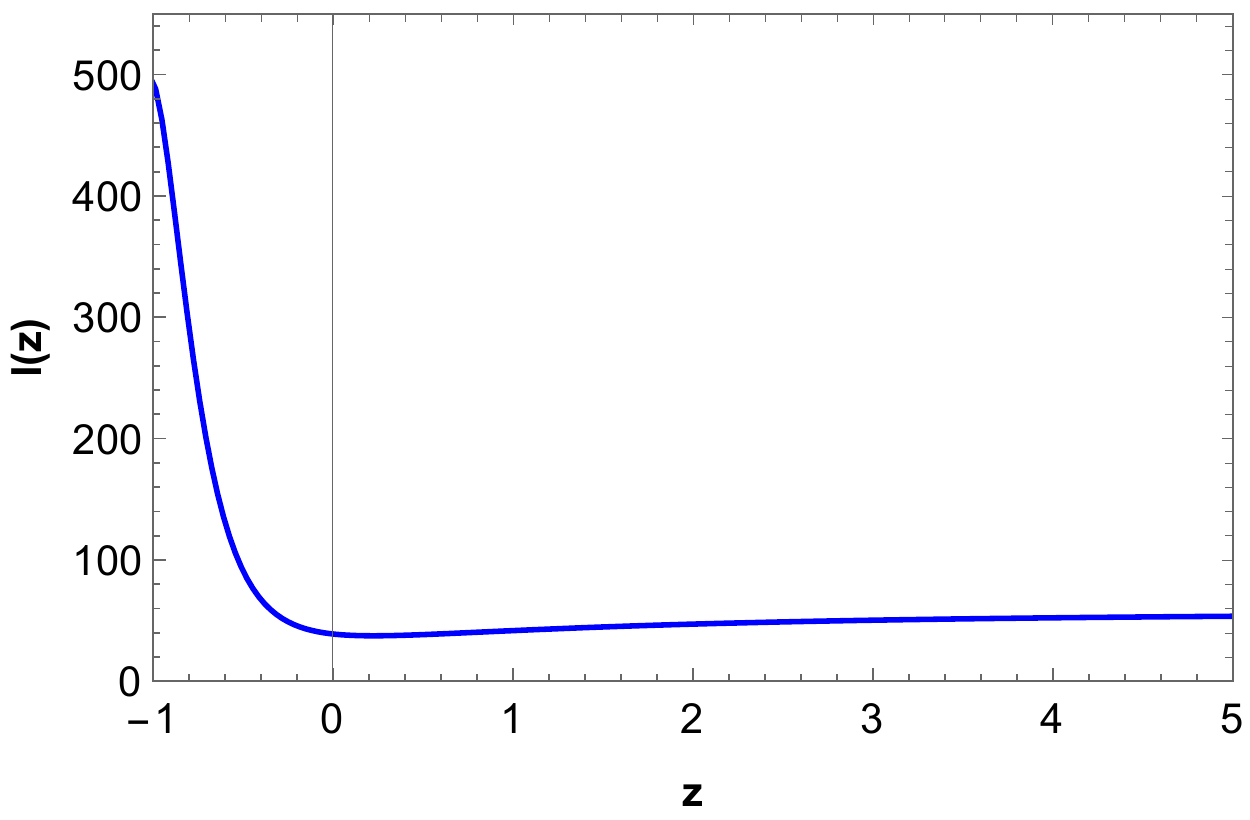}
    \caption{The plots are the geometrical parameters for model M1. The plots are of deceleration parameter (q), jerk parameter (j), snap parameter (s), and lerk parameter (l) respectively.}
    \label{M1p}
\end{figure}
\end{widetext}
The first plot in figure (\ref{M1p}) shows the deceleration parameter (q). Which is the very important quantity that describes the acceleration of the universe. In the plot at present time ($z=0$), the rate of deceleration is $q_0=-0.54$ which is a clear match of the model with the current phase of the accelerating universe. Another important quantity we can find from the plot is that the universe had a phase transition at $z_{tr}=0.71$ which is the point when $q=0$, the universe stopped decelerating and started accelerating and will continue accelerating from the prediction of the model.\\ 
The higher-order derivatives of the deceleration parameters, known as the jerk (j), snap (s), and lerk (l) parameters are plotted along with the deceleration parameter in (\ref{M1p}). These parameters are useful in the way to show the evolution of the universe from past to future. The sign of the jerk parameter controls the change in the late-time universe's dynamics, a positive value indicating the happening of a transition time during which the universe modifies its expansion. As well, the value of the snap parameter is important to discriminate an evolving dark energy term or a cosmological behavior. Similar arguments are discussed in  \cite{Capozziello:2019cav, Aviles:2012ay}. For model M1 we obtained the values, $j_0=1.21$, $s_0=1.37$, and $l_0=38.98$.

\subsection{Evolution of Cosmographic Parameters in model M2}

Similarly, we get geographical parameters associated with model M2 as follows.
\begin{equation}
    q(z)=-1+\alpha\left[1-\frac{3}{({1+(\beta(1+z))^{2 \alpha}})}\right]
\end{equation}

\begin{widetext}
\small
\begin{equation}
    j(z)=1+3\alpha\left[-1+\frac{3}{({1+(\beta(1+z))^{2 \alpha}})}\right]+\alpha^2\left[2+\frac{12}{({1+(\beta(1+z))^{2 \alpha}})^2}-\frac{6}{({1+(\beta(1+z))^{2 \alpha}})}\right]
\end{equation}
\begin{multline}
    s(z)=1+\alpha\left[-6+\frac{18}{({1+(\beta(1+z))^{2 \alpha}})}\right]+\alpha^2\left[11-\frac{42}{({1+(\beta(1+z))^{2 \alpha}})}+\frac{75}{({1+(\beta(1+z))^{2 \alpha}})^2}\right]\\+\alpha^3\left[-6+\frac{24}{({1+(\beta(1+z))^{2 \alpha}})}-\frac{30}{({1+(\beta(1+z))^{2 \alpha}})^2}+\frac{60}{({1+(\beta(1+z))^{2 \alpha}})^3}\right] 
\end{multline}
\begin{multline}
    l(z)=1+\alpha\left[-10+\frac{30}{({1+(\beta(1+z))^{2 \alpha}})}\right]+\alpha^2\left[35-\frac{150}{({1+(\beta(1+z))^{2 \alpha}})}+\frac{255}{({1+(\beta(1+z))^{2 \alpha}})^2}\right]\\+\alpha^3\left[-50+\frac{240}{({1+(\beta(1+z))^{2 \alpha}})}-\frac{450}{({1+(\beta(1+z))^{2 \alpha}})^2}+\frac{660}{({1+(\beta(1+z))^{2 \alpha}})^3}\right]\\+\alpha^4\left[24-\frac{120}{({1+(\beta(1+z))^{2 \alpha}})}+\frac{240}{({1+(\beta(1+z))^{2 \alpha}})^2}-\frac{120}{({1+(\beta(1+z))^{2 \alpha}})^3}+\frac{360}{({1+(\beta(1+z))^{2 \alpha}})^4}\right]
\end{multline}
\begin{figure}[htbp]
    \centering
    \includegraphics[width=0.36\linewidth]{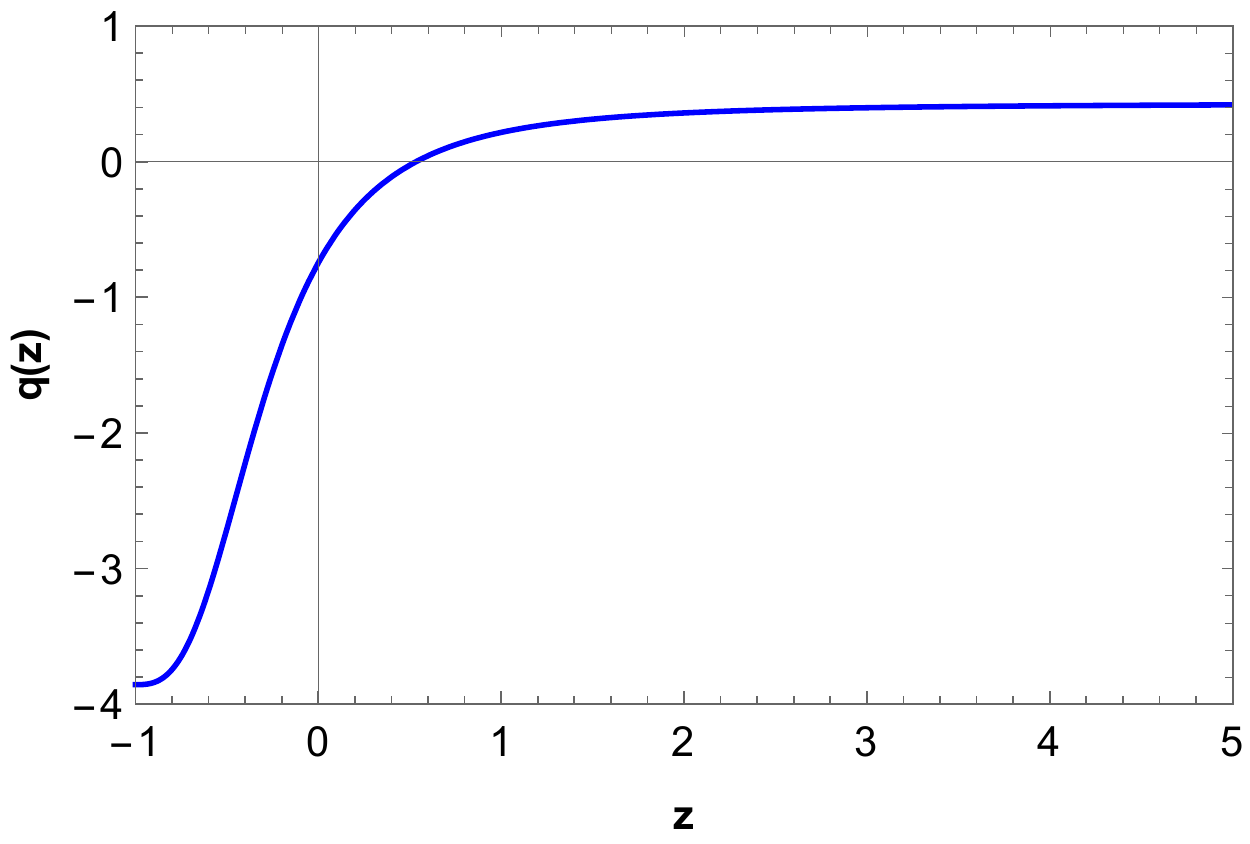}
    \includegraphics[width=0.36\linewidth]{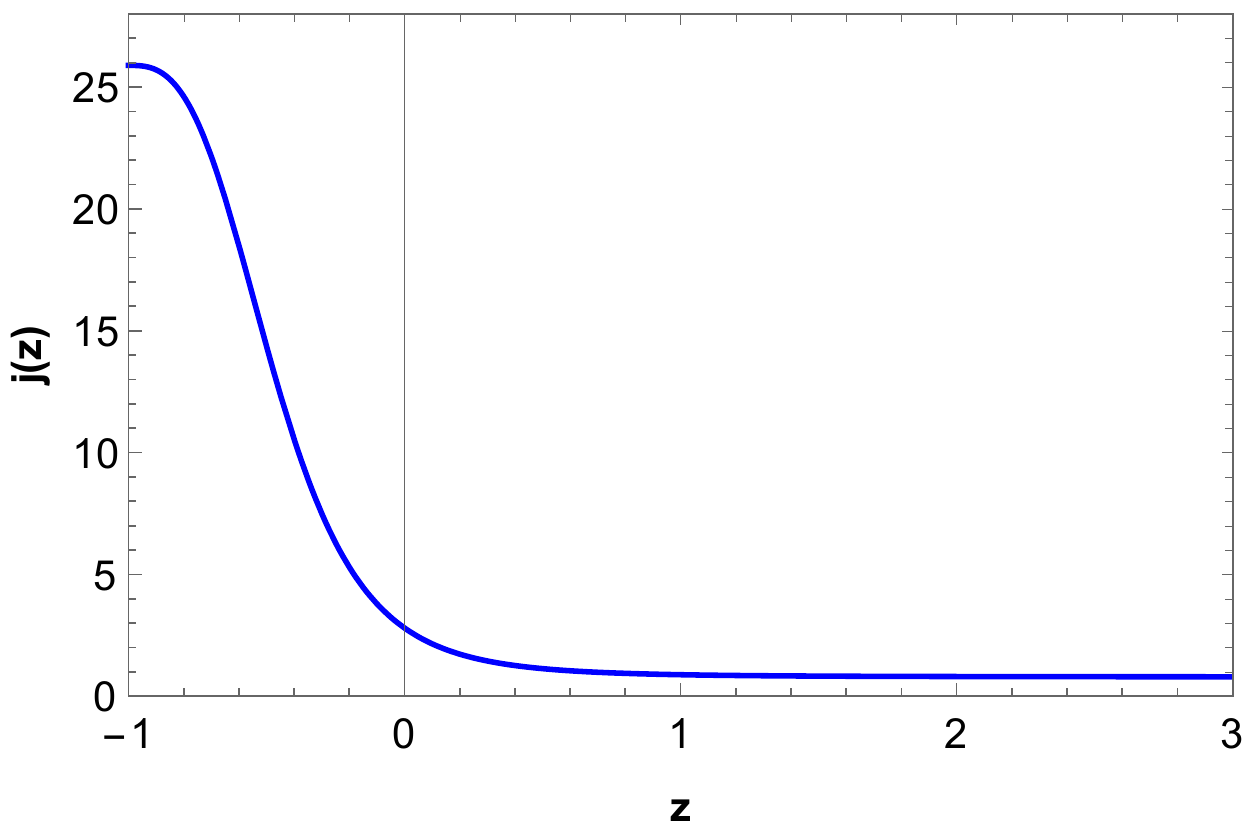}
    \includegraphics[width=0.36\linewidth]{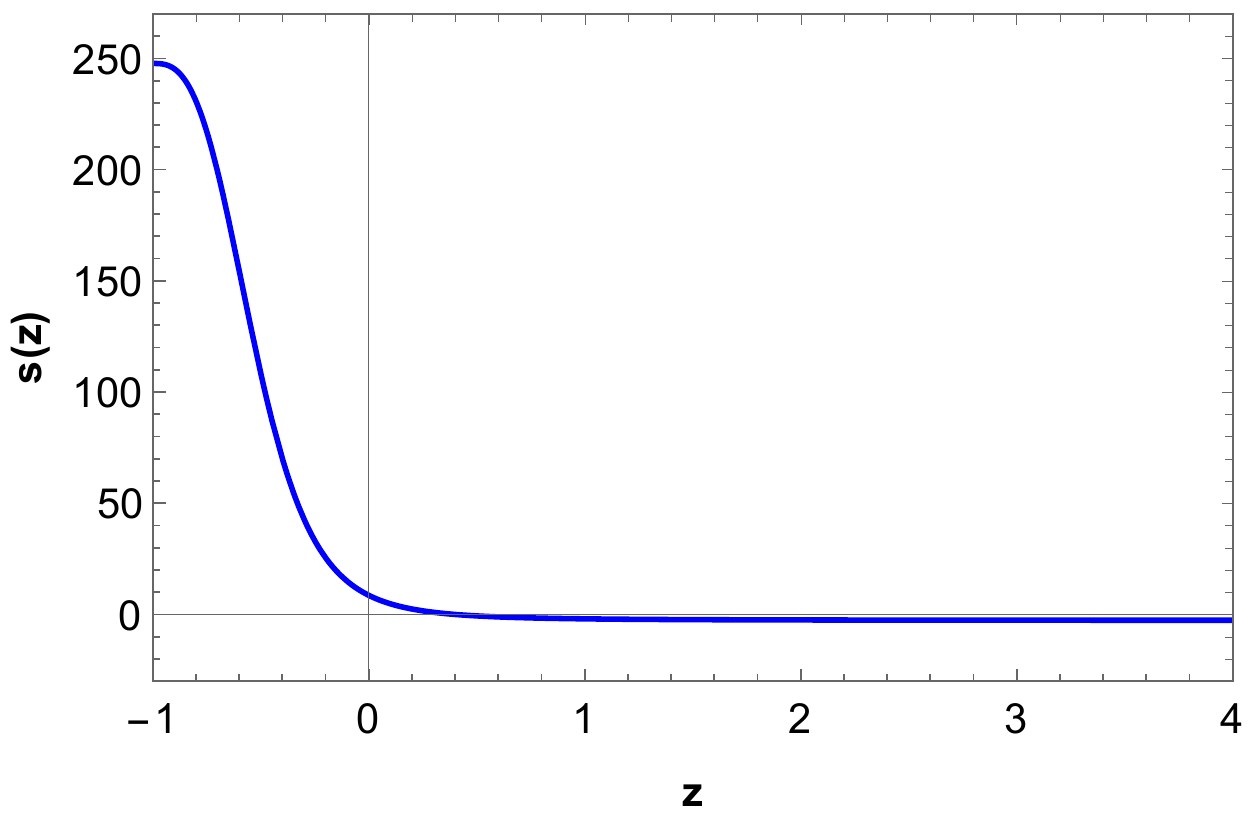}
    \includegraphics[width=0.36\linewidth]{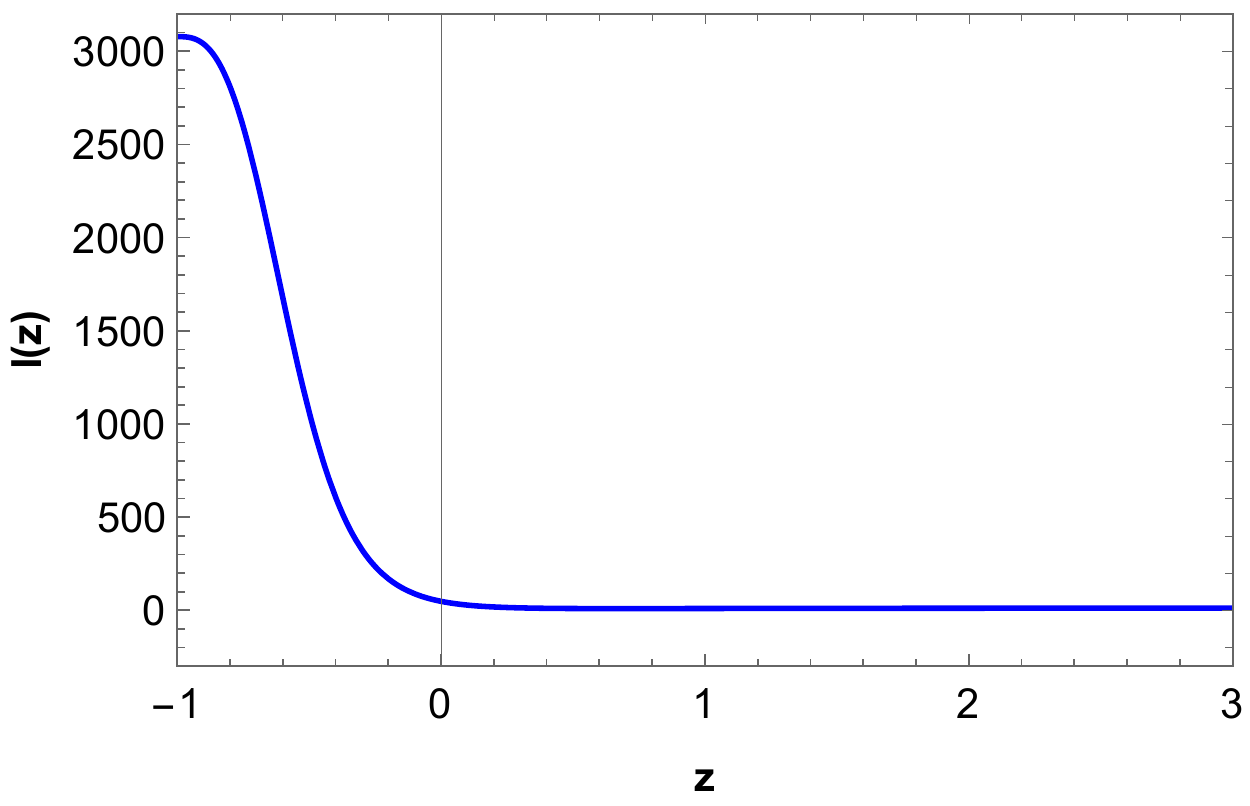}
    \caption{The plots are the geometrical parameters for model M2. The plots are of deceleration parameter (q), jerk parameter (j), snap parameter (s), and lerk parameter (l) respectively.}
    \label{M2p}
\end{figure}
\end{widetext}
In figure (\ref{M2p}) we find the behavior of model M2 similar to model M1 for the deceleration parameter. The transition phase occurs for M2 at $z_{tr}=0.54$. The value of other parameters in present ($z=0$) are $q_0=-0.77$, $j_0=2.6$, $s_0=7.87$, and $l_0=48.35$. 

In the next section, We discuss some kinematic tests for the considered models for the validation of the models.

\section{Kinematic Tests}

In this section, we are going to discuss some kinematic behavior of the models. As the distance between two comoving objects for an Earth-bound observer constantly changes due to the accelerated expansion of the universe, there are some kinematic tests to characterize the distance measurements of the universe, such as lookback time, proper distance, luminosity distance, and comoving volume are discussed in the following subsections. 

\subsection{Lookback Time}

We know that light travels at a constant finite speed and it takes time to reach an object at a far distance while also considering the redshift. Hence observing space is always a looking back in time. The lookback time ($t_L$) is the time difference between the age of the universe from the observations and the age of the universe when the photons were emitted. If the photon was emitted by a star at any instance of time $t$ and it reaches Earth at the time $t_0$ then the lookback time is defined by \cite[pp.~313-315]{Peebles:1994xt}
\begin{equation}
    t_L = t_0 - t = \int_{a_0}^a \frac{da}{\Dot{a}}
\end{equation}
Where $a_0$ is the value of the scale factor at present and $\Dot{a}$ is the time derivative of the scale factor. 

We have already derived the $t-z$ relation for models M1 and M2 in Eqs. (\ref{M1z}) and (\ref{M2z}) which can be rewritten as, 
for M1
\begin{equation}
    t(z)= \frac{1}{H_0}\frac{(1+\beta^\alpha)^2}{\alpha \beta^\alpha \big(1+\big(\beta(1+z)\big)^\alpha\big)}
    \label{M1z1}
\end{equation}
for M2
\begin{equation}
    t(z)= \frac{1}{H_0}\frac{(1+\beta^{2\alpha})^{3/2}}{\alpha \beta^{2\alpha} \big(1+\big(\beta(1+z)\big)^{2\alpha}\big)^{1/2}}
    \label{M2z1}
\end{equation}
With the help of these $ t-z$ relations, we can establish the functional form of lookback time.
for M1
\begin{equation}
\small
    t_L(z)= \frac{(1+\beta^\alpha)}{\alpha H_0 \beta^\alpha}-\frac{(1+\beta^\alpha)^2}{\alpha H_0 \beta^\alpha \big(1+\big(\beta(1+z)\big)^\alpha\big)}
\end{equation}
for M2
\begin{equation}
\small
    t_L(z)= \frac{(1+\beta^{2\alpha})}{\alpha H_0 \beta^{2\alpha}}-\frac{(1+\beta^{2\alpha})^{3/2}}{\alpha H_0 \beta^{2\alpha} \big(1+\big(\beta(1+z)\big)^{2\alpha}\big)^{1/2}}
\end{equation}

 As we have considered $H_0=69 {km/s}/{Mpc}$, by multiplying the factor of 
 \begin{equation*}
     \frac{{3.0857 \times 10^{19} km/Mpc}}{{3.1536 \times 10^7 s/year}}
 \end{equation*}
 with the function of $t_L (z)$ we get the following plot of lookback time in years for redshift. 

\begin{figure}[htbp]
    \centering
    \includegraphics[width=1\linewidth]{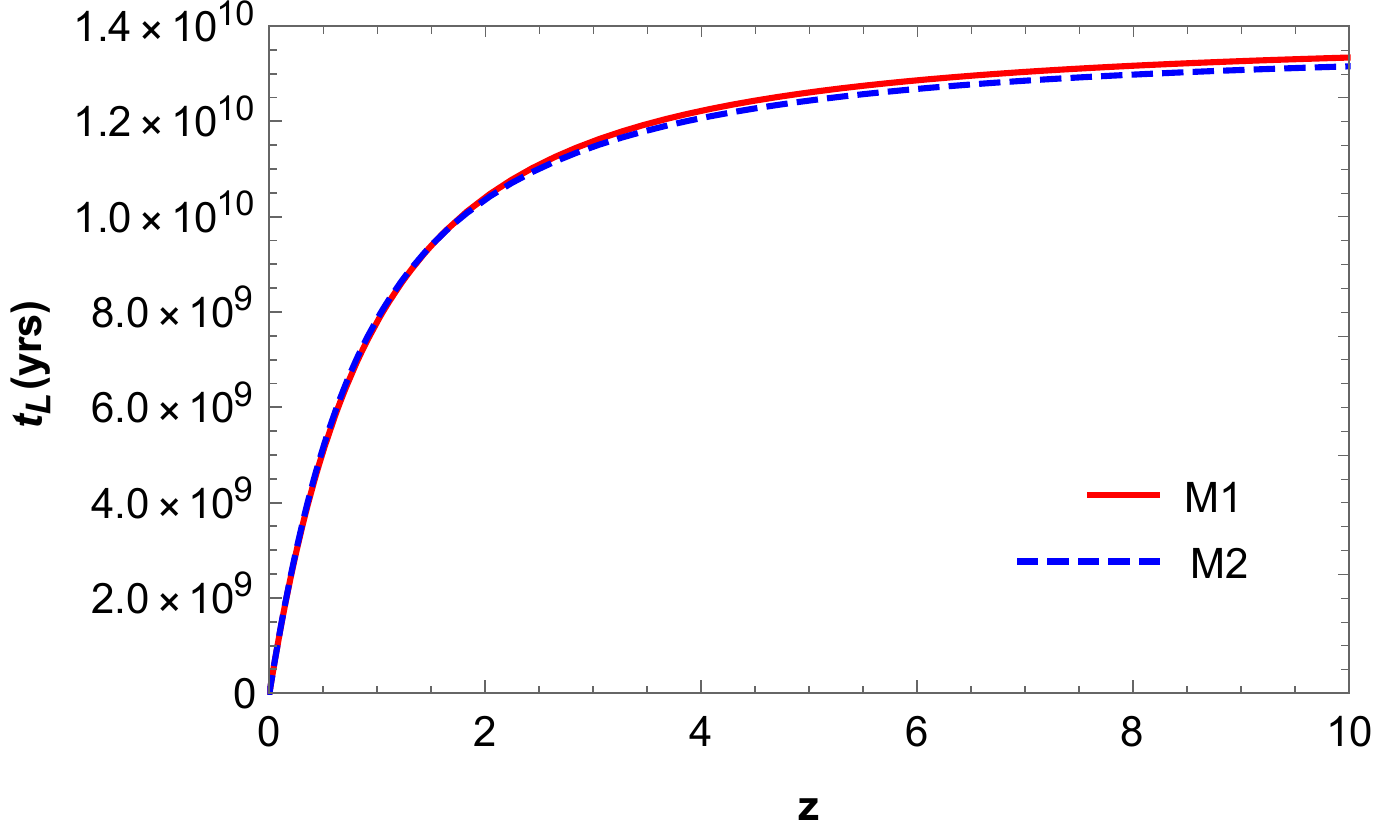}
    \caption{Lookback time in terms of redshift}
    \label{fig.lookback}
\end{figure}
In Fig. (\ref{fig.lookback}) we find that the lookback time becomes very crucial when we measure the large distance. As we can see the lookback time is in order of billion years for a considerable amount of redshift. We also observe that for models M1 and M2, the lookback time plot is almost similar.\\
We can also find the age of the universe from Eqs. (\ref{M1z1}) and (\ref{M2z1}) by imputing $z=0$ for models M1 and M2 respectively. The current age of the universe is $13.8039$ billion years according to model M1 and $13.6766$ billion years according to model M2.

\subsection{Proper Distance}

The light requires time to travel from the source to the observer, the distance between the source and the observer at the time when the light was emitted can be found from the simple formula $d_p(z)=a_0 r(z)$ where $r(z)$ is the radial distance of the object. The term is called the proper distance. 

\begin{equation}
    r(z)=\int_t^{t_0} \frac{dt}{a(t)} 
\end{equation}
for M1
\begin{equation}
\small
    d_p(z)= \frac{(t_0 -t)}{\beta} \bigg[ \bigg(\frac{(1+\beta^\alpha)^2}{H_0 \alpha \beta^\alpha}\bigg)^{1/\alpha} \frac{\alpha(t_0 -t)^{-1/\alpha}}{\alpha-1} -1 \bigg]
\end{equation}

for M2
\begin{equation}
\small
    d_p(z)= \frac{(t_0 -t)}{\beta} \bigg[ \bigg(\frac{(1+\beta^{2\alpha})^{3/2}}{H_0 \alpha \beta^{2\alpha}}\bigg)^{1/\alpha} \frac{\alpha(t_0 -t)^{-1/\alpha}}{\alpha-1} -1 \bigg]
\end{equation}
To have the value in light-years (ly) we multiply the proper distance function form with the factor of
\begin{equation*}
    \frac{{3.0857 \times 10^{19} km/Mpc}\times 2.9972 \times 10^5 km/s}{{9.4607 \times 10^{12} km/ly}}
\end{equation*}
The plot of proper distance (in light years) versus redshift is then plotted in the next figure for models M1 and M2 as red and dashed blue lines respectively. 

\begin{figure}[htbp]
    \centering
    \includegraphics[width=1\linewidth]{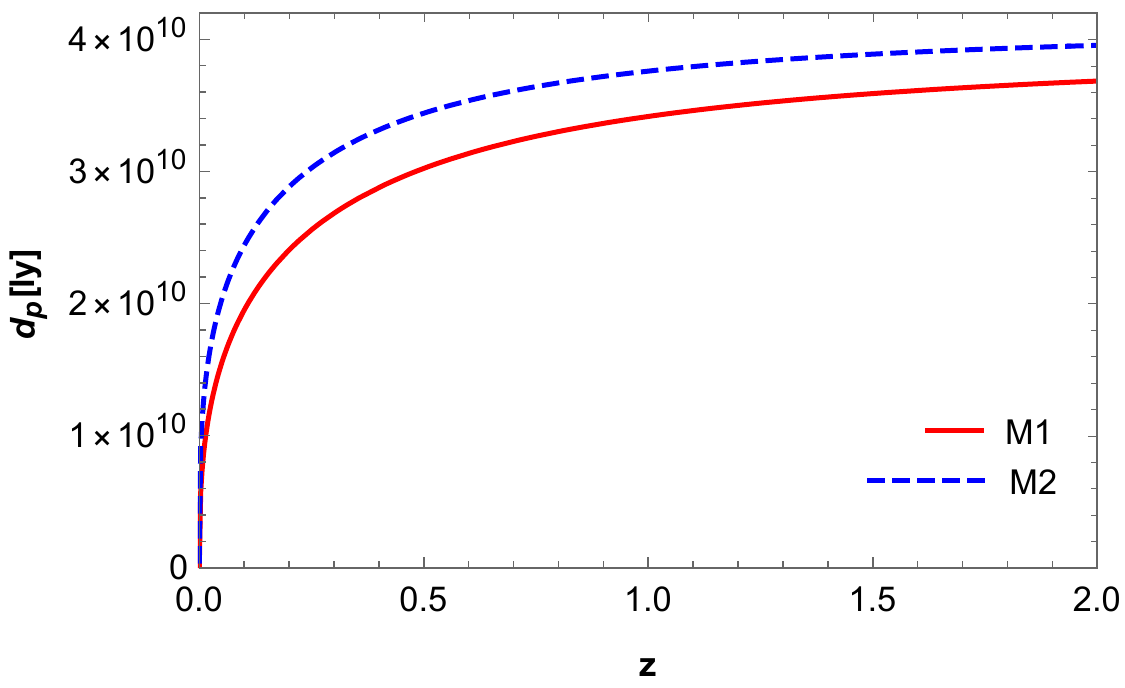}
    \caption{Proper distance in terms of redshift}
    \label{fig.ProperDistance}
\end{figure}

from Fig (\ref{fig.ProperDistance}) we find that, as we observe the larger redshift, the proper distance between the object and the observer is larger. However, after a certain redshift value, the distance gets nearly saturated.

\subsection{Luminosity Distance}

The luminosity distance is one of the best ways to unfold the trace of the evolution of the universe. The luminosity distance can be defined as $d_l= \big(\frac{l}{4\pi L} \big)^{1/2}$, where, L is the apparent luminosity flux and l is the luminosity of the object \cite[pp.~420-424]{Weinberg:1972kfs}. The luminosity density in terms of redshift in FLRW cosmology can be given as 
\begin{equation}
    d_l = a_0 (1+z) r(z) = (1+z) d(z)
\end{equation}
Hence for model M1
\begin{equation}
    \small
    d_l(z)= \frac{(t_0 -t)}{\beta(1+z)^-{1}} \bigg[ \bigg(\frac{(1+\beta^\alpha)^2}{H_0 \alpha \beta^\alpha}\bigg)^{1/\alpha} \frac{\alpha(t_0 -t)^{-1/\alpha}}{\alpha-1} -1 \bigg]
\end{equation}
for M2
\begin{equation}
\small
    d_l(z)= \frac{(t_0 -t)}{\beta(1+z)^{-1}} \bigg[ \bigg(\frac{(1+\beta^{2\alpha})^{3/2}}{H_0 \alpha \beta^{2\alpha}}\bigg)^{1/\alpha} \frac{\alpha(t_0 -t)^{-1/\alpha}}{\alpha-1} -1 \bigg]
\end{equation}

\begin{figure}[htbp]
    \centering
    \includegraphics[width=1\linewidth]{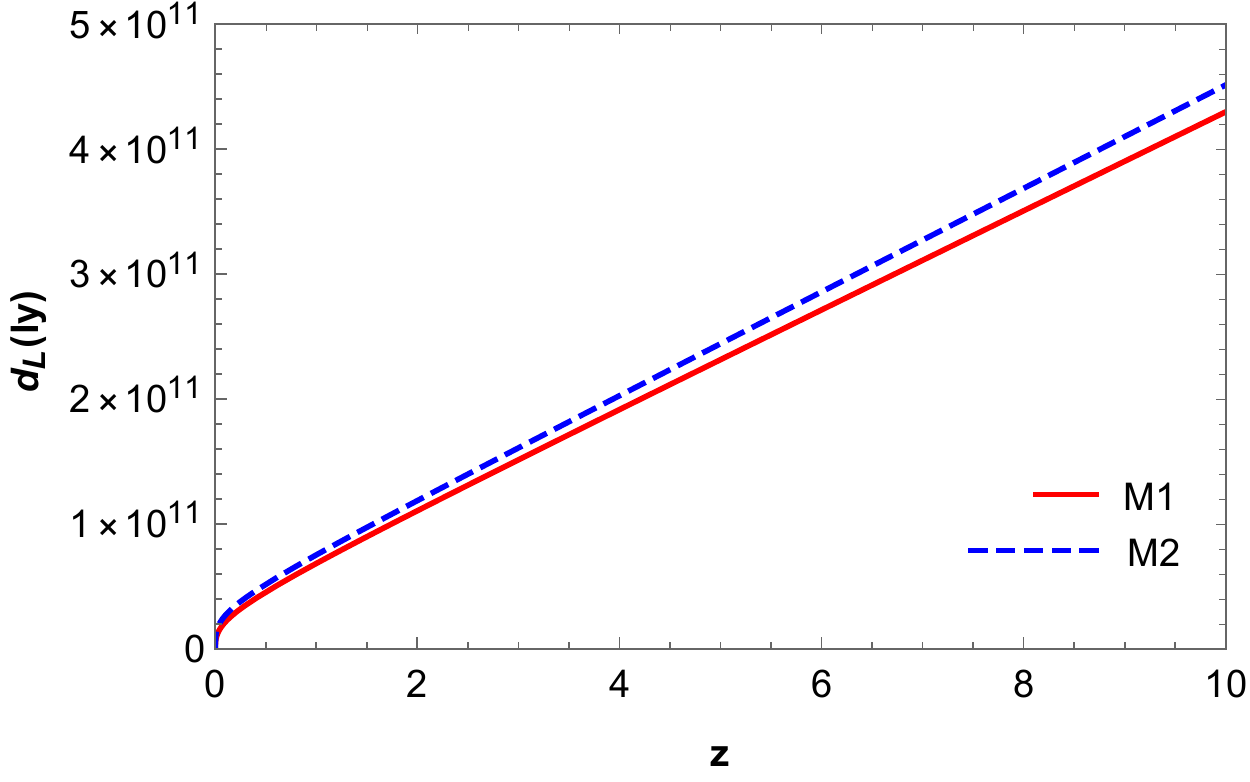}
    \caption{Luminosity distance in terms of redshift}
    \label{fig.LuminosityDistance}
\end{figure}

In Fig. (\ref{fig.LuminosityDistance}) we observe the luminosity distance plot is linear in terms of redshift. for a larger redshift, the distance is larger. 

\subsection{Angular Diameter Distance}

The ratio of the physical transverse size of an object to its angular size is called the angular diameter distance $d_A$ \cite[pp.~325-327]{Peebles:1994xt}. It can be written as 

\begin{equation}
    d_A = \frac{d_p(z)}{1+z} = \frac{d_l}{(1+z)^2}
\end{equation}
Deriving the function for models, for M1
\begin{equation}
    \small
    d_A(z)= \frac{(t_0 -t)}{\beta(1+z)} \bigg[ \bigg(\frac{(1+\beta^\alpha)^2}{H_0 \alpha \beta^\alpha}\bigg)^{1/\alpha} \frac{\alpha(t_0 -t)^{-1/\alpha}}{\alpha-1} -1 \bigg]
\end{equation}
for M2
\begin{equation}
\small
    d_A(z)= \frac{(t_0 -t)}{\beta(1+z)} \bigg[ \bigg(\frac{(1+\beta^{2\alpha})^{3/2}}{H_0 \alpha \beta^{2\alpha}}\bigg)^{1/\alpha} \frac{\alpha(t_0 -t)^{-1/\alpha}}{\alpha-1} -1 \bigg]
\end{equation}

\begin{figure}[htbp]
    \centering
    \includegraphics[width=1\linewidth]{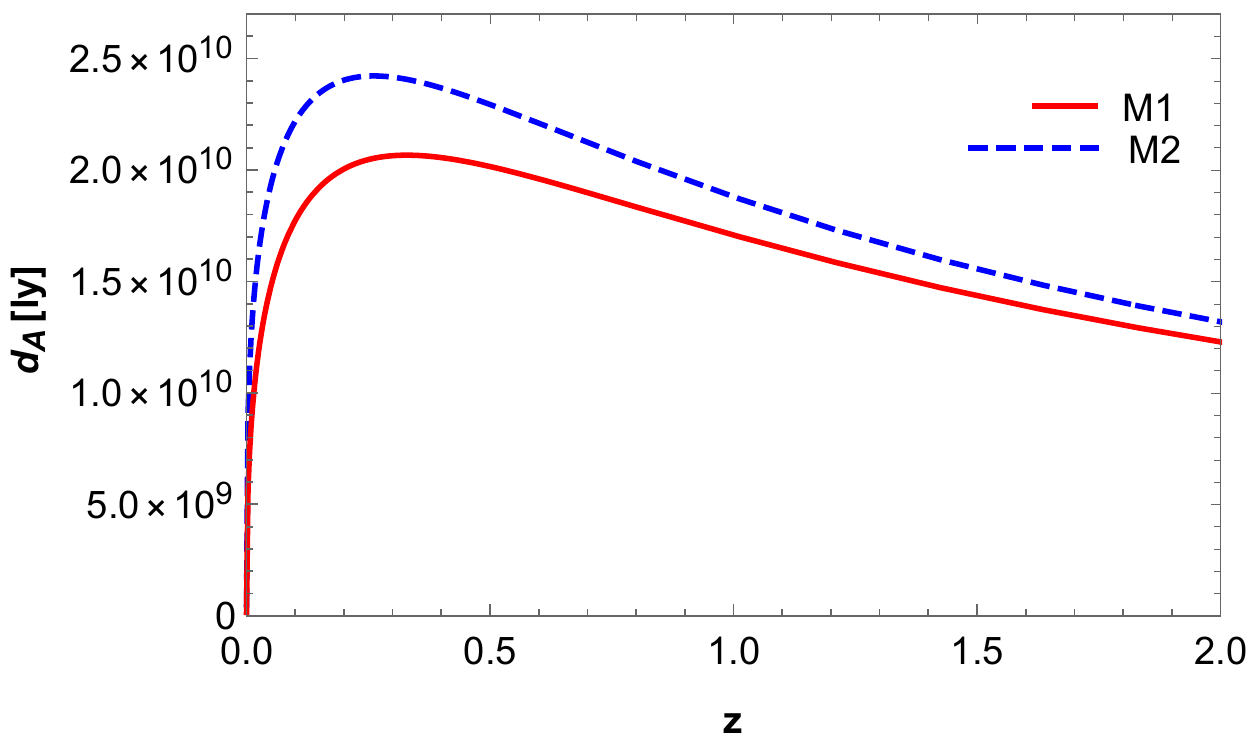}
    \caption{Angular diameter distance in terms of redshift}
    \label{fig.AngularDistance}
\end{figure}
According to Fig. (\ref{fig.AngularDistance}) the angular diameter distance increases initially which starts falling eventually after a certain peak in the distance. 

\begin{table}[htbp]
    \centering
    \setlength{\tabcolsep}{2pt}
    \renewcommand{\arraystretch}{1.2}
    \begin{tabular}{|c|c|c|c|c|c|}
    \hline
    \multicolumn{6}{|c|}{Distance measurements}
    \\ 
    \hline
    \hline
    Redshift & Model & $t_L$ ($10^9$yrs) & $d_p$ ($10^9$ly) & $d_l$ ($10^9$ly) & $d_A$ ($10^9$ly) \\
    \hline
    \multirow{2}{*}{0.5} & M1 & 5.0959 & 30.229 & 45.343 & 20.153 \\
                         & M2 & 5.1713 & 34.406 & 51.609 & 22.927 \\
    \hline
    \multirow{2}{*}{1.0} & M1 & 7.8019 & 34.159 & 68.318 & 17.079 \\
                         & M2 & 7.8618 & 37.591 & 75.183 & 18.796 \\
    \hline
    \multirow{2}{*}{5.0} & M1 & 12.607 & 38.580 & 231.48 & 6.4299 \\
                         & M2 & 12.436 & 40.707 & 244.24 & 6.7845 \\
    \hline
    \multirow{2}{*}{10} & M1 & 13.339 & 39.079 & 429.87 & 3.5527 \\
                         & M2 & 13.154 & 41.038 & 451.41 & 3.7307 \\
    \hline
\end{tabular}
    \caption{Kinematic tests for different redshift}
    \label{tab.kinematic tests}
\end{table}

\section{Condition of Stability}

The speed of sound could be a key to describing the scale of gravitational instability. The square sound speed ($c_s^{\;2}$) must be less than the speed of light in the universe, hence it must be $c_s^{\;2} \in [0,1]$. 
The square of the speed of sound can be defined as
\begin{equation}
    c_s^{\;2}=\frac{dp}{d\rho}=\frac{{dp}/{dz}}{{d\rho}/{dz}}
\end{equation}
We find  $c_s^{\;2}$ for model M1 and M2 as following \\
for M1
\begin{multline}
        c_s^{\;2}= \bigg[\big((1+z)\beta \big)^{2\alpha}\Big(8\pi(2\alpha-3) +3\nu(\alpha-1)\Big)
        \\
       +3\nu (1+\alpha)-\alpha \big((1+z)\beta \big)^\alpha (16\pi +3\nu) 
       \\
        +8\pi (3+2\alpha) \bigg]/\bigg[\big((1+z)\beta \big)^{2\alpha}\Big(24\pi + (3+\alpha)\nu \Big)
        \\
        -3 (8\pi +\nu)+\alpha\nu\Big(1- \big((1+z)\beta \big)^\alpha \Big)\bigg] 
\end{multline}
for M2
\begin{multline}
\small
        c_s^{\;2}= \bigg[\big((1+z)\beta \big)^{2\alpha}\Big(8\pi (3-2\alpha)+3\nu(1-\alpha)\Big)
        \\
        \cdot \Big(1-\big((1+z)\beta \big)^{2\alpha}\Big)+16\pi (3+4\alpha) 
        \\
        + 6\nu (1+2\alpha)\bigg] / \bigg[-48\pi -2\nu (3-2\alpha)
        \\
        -\big((1+z)\beta \big)^{2\alpha}\Big(24\pi +\nu(3+\alpha)\Big)
        \\
        \cdot \Big(1-\big((1+z)\beta \big)^{2\alpha}\Big) \bigg] 
\end{multline}
\begin{figure}[htbp]
        \centering
        \includegraphics[width=1\linewidth]{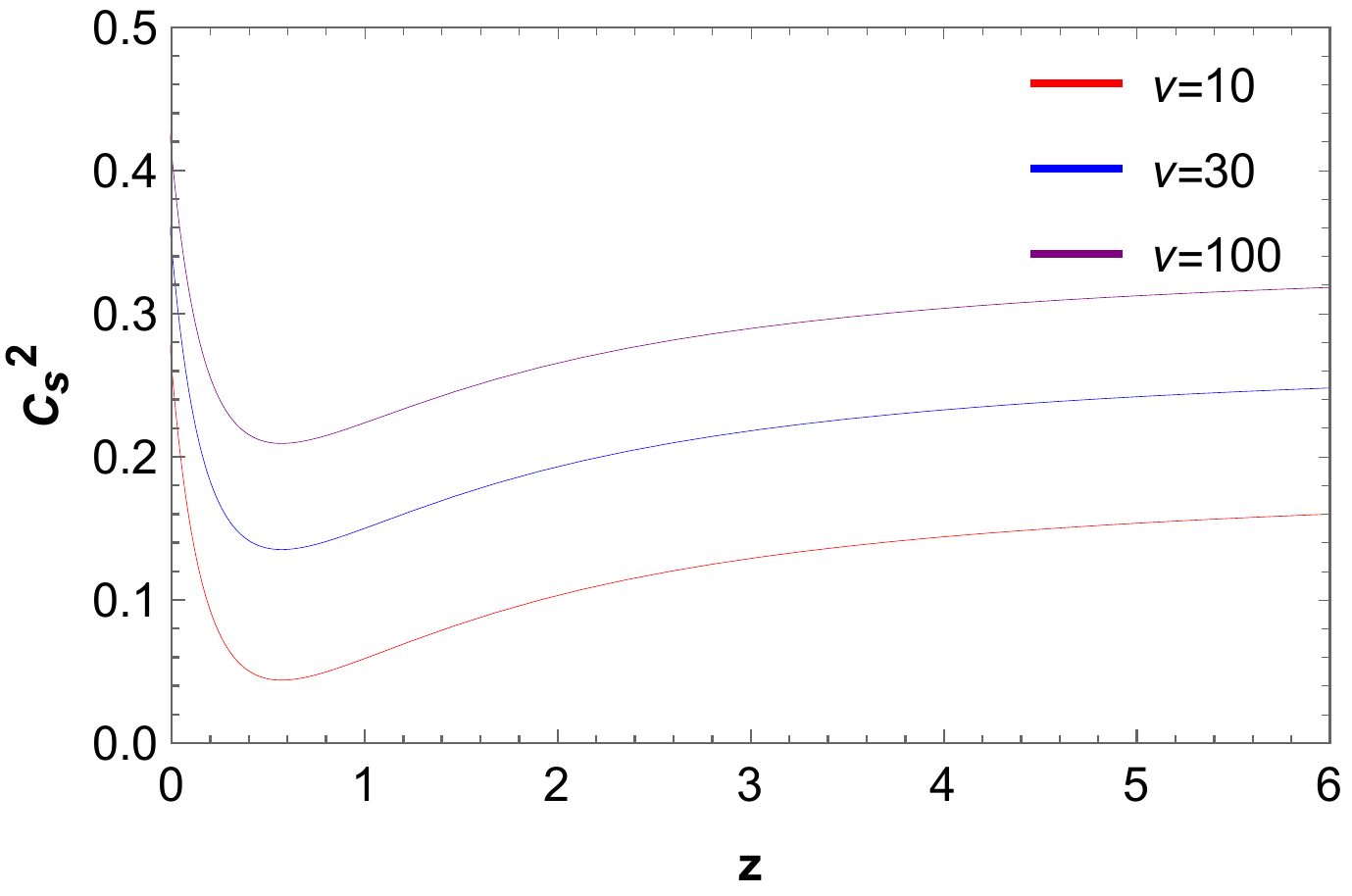}
        \caption{$c_s^{\;2}$ in terms of redshift for model M1}
        \label{fig.M1sos}
\end{figure}
\begin{figure}[htbp]
        \centering
        \includegraphics[width=0.95\linewidth]{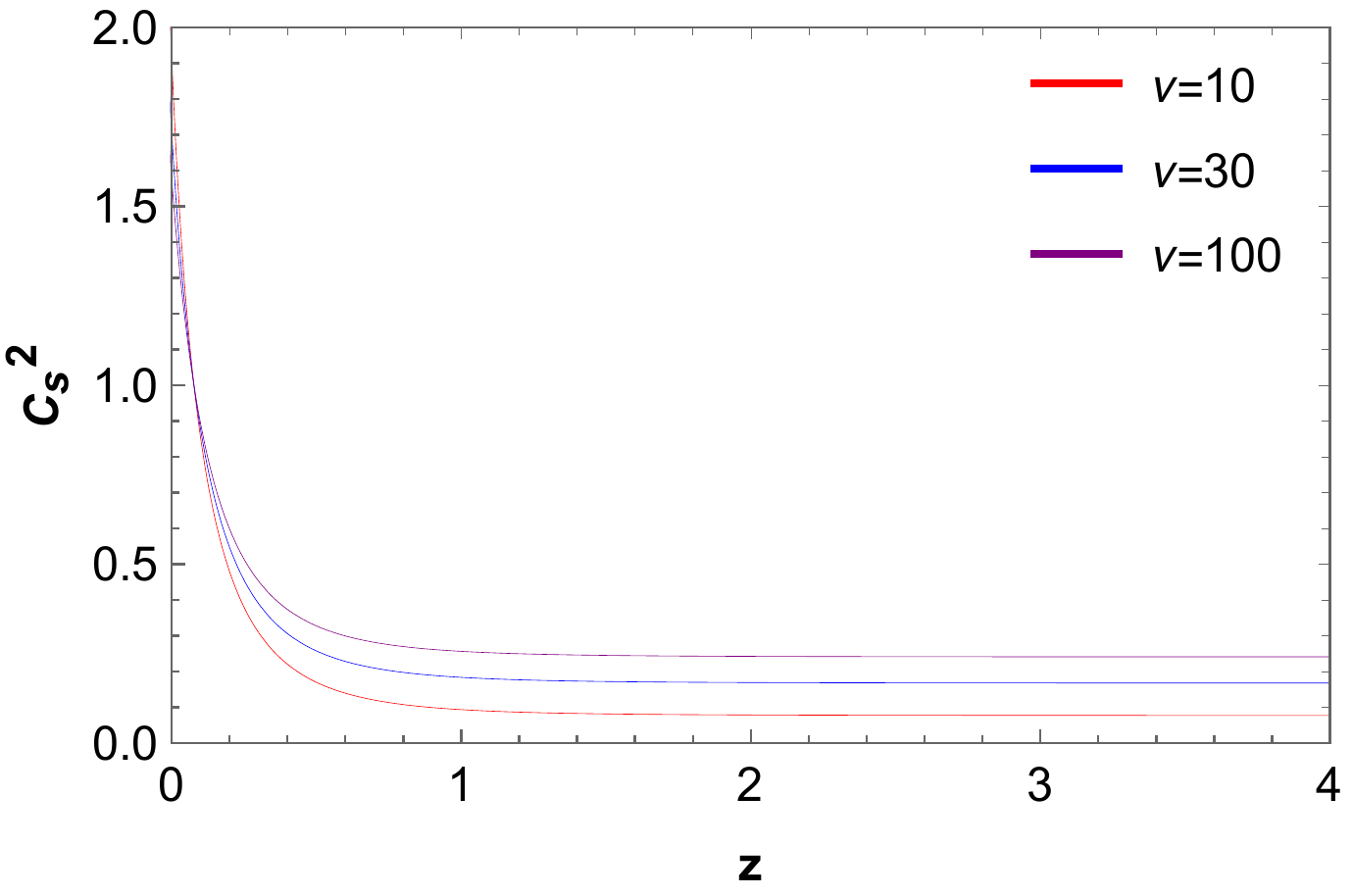}
        \caption{$c_s^{\;2}$ in terms of redshift for model M2}
        \label{fig.M2sos}
\end{figure}

The plot of $c_s^{\;2}$ in Figs. (\ref{fig.M1sos}) and (\ref{fig.M2sos}) are of model M1 and M2 respectively. In the plot of M1, we find out that the value of the square of the speed of the sound is $0<c_s^{\;2}<1$ hence the model M1 is stable. However, in model M2 the value of $c_s^{\;2}$ is slightly larger than 1 showing the possibility of an unstable universe. 

\section{Concluding remarks}

In the current research, we tried to investigate the evolution of the universe under the consideration of the cosmological model: parametrization of Hubble by performing some of the cosmological tests in the extended version of symmetric teleparallel gravity $f(Q,T)$. 

Using the linear functional form $f(Q,T)=\mu Q+ \nu T$ in the gravity action, and barotropic EoS, we find the value of EoS parameter at present ($z=0$) when $\nu=100$ is $\omega_0=-0.5058$ and $\omega_0=-0.7878$ for model M1 and M2 respectively showing the quintessence phase. Shortly after in near future, the EoS parameter will have a value below -1 predicting the phantom era for both models. Besides, to have a positive value of $\omega$ at large z, $\nu$ must be greater than $-2.6$ for model M1 and $\nu >2.81$ for model M2. 

Next, we attempt to disclose a crucial quantity, the validity of the energy conditions. Hence the energy density equation was obtained and found to be always on the positive side. However, the strong energy condition gets violated for the current phase of the universe promoting the accelerated expansion in models M1 and M2. On the other hand, the dominant energy condition and null energy condition are true at $z=0$ however, it does not hold in the near future. Hence we plot the variation of model parameters plot. The boundary condition for which the above description is true is when $\mu<0$ and $\nu>-12.56$. 

With the use of the model-independent approach, we get to know certain features of the model without any assumptions other than the flatness of the universe. To confine its values with the current observations, we find the geometrical parameter values at $z=0$ to compare them with present values. which are shown in the following table. 

\begin{table}[htbp]
  \begin{center}
    \begin{tabular}{|l|c|c|} 
    \hline
    \multicolumn{3}{|c|}{Values of Cosmographic Parameters at $z=0$ (at present)}
    \\ 
    \hline
    \hline
      \textbf{Parameters} & \textbf{Model M1} & \textbf{Model M2}\\
      \hline
      Deceleration parameter ($q_0$) &-0.5372 &-0.7462 \\
      Jerk parameter ($j_0$)         &1.1996  &2.8025 \\
      Snap parameter ($s_0$)         &1.4032  &8.6445 \\
      Lerk parameter ($l_0$)         &38.9871  &48.3503\\
      \hline
    \end{tabular}
  \end{center}
  \caption{Cosmographic parameters at $z=0$}
  \label{tab:table1}
\end{table}
In the table (\ref{tab:table1}) we find the geometrical parameter values at $z=0$ for models M1 and M2 respectively. The value of the deceleration parameter ($q_0$) confirms the observational evidences of the accelerated phase of the universe expansion for both models. We also find at $z>0.7364$ for model M1 and $z>0.5350$ for model M2 the deceleration parameter is positive suggesting the decelerated expansion of the universe. The point when the deceleration phase changes into the accelerating phase is called the transition. The positive value jerk parameter supports the accelerated expansion evidence. Interestingly, for $\Lambda CDM$ the value should always be $j_0=1$ which contradicts our current observations. The following geometrical parameters, snap and lerk parameter gives the dynamics of the universe. We extended our model-independent approach analysis to obtain some of the kinematic behavior of the models. We found lookback time, proper distance, luminosity distance, and angular diameter distance. 

In the end, we try to find the stability of the universe. The square of the speed of sound must be less than the squared speed of light. Hence the condition $c_s^{\;2} \in [0,1]$ must hold to have a stable universe. For model M1 we find the stable universe solution however, for model M2 the slightly larger than 1 value of $c_s^{\;2}$ suggests the possibility of the unstable universe solution.

Although GR made precise predictions to describe the cosmological phenomena, it still lacks to explain some of the undesirable effects in the dark sectors to match the consistency of GR with the observational data hence, the approach of the modified theory of gravity. In that regard, we performed various cosmological tests to solve some of the most fundamental problems with the universe in the $\Lambda$ cold dark matter model. It seemed to have a very good approximation of the evolution of the universe from the current study in this paper. However, we still remained an open issue of the modification in the geometry of the gravity as GR still holds pretty well in almost every circumstance. Hence, more research must be done on the $f(Q,T)$ gravity.

\textbf{Acknowledgement: } Author SKJP thanks IUCAA, Pune for hospitality and
other facilities under its IUCAA associateship program, where a large part of
work has been done.

\textbf{Author contributions:} The calculations, plotting, manuscript writing, and overall manuscript preparation is done by VK under the supervision of SKJP, who has administered the project and completed the project. The manuscript has been read and approved by all authors. 

\textbf{Funding:} There is no fund available for the publication of this
research article.

\textbf{Data Availability Statement:} This manuscript has not used any data for the work.

\textbf{Conflict of interest:} The authors have no relevant financial or
non-financial interests to disclose.

\textbf{Ethical statements:} The submitted work is original and has not been
published anywhere else.



\end{document}